\documentclass{article}
\usepackage{jheppub} 
\usepackage[utf8x]{inputenc}	
\usepackage[english]{babel}

\usepackage[ddmmyyyy,hhmmss]{datetime}

\allowdisplaybreaks
\usepackage{amsxtra}
\usepackage{mathtools}
\usepackage{float}
\usepackage{tikz}
\usepackage{amsmath}
\usepackage{graphicx}
\usepackage{makecell}

\usepackage{enumitem}

\usepackage{todonotes}
\usepackage[normalem]{ulem}

\newcommand{\ii}{\mathrm{i}} 
\newcommand{\dd}{\mathrm{d}}
\newcommand{\ee}{\mathrm{e}}

\makeatletter
\def\@fpheader{}
\makeatother

\newtheorem{theorem}{Theorem}[section]
\newtheorem{definition}{Definition}[section]

\title{5d Higgs branch and instanton magnetization}

\preprint{Imperial/TP/26/AH/03}

\author[1]{Amihay Hanany}
\author[2]{Alessandro Tomasiello}
\author[1]{Elias Van den Driessche}

\affiliation[1]{Abdus Salam Centre for Theoretical Physics,\\ Imperial College London,\\ Prince Consort Road, SW7 2AZ, UK}
\affiliation[2]{Dipartimento di Matematica,\\ Università degli Studi di Milano-Bicocca, \\
Via Cozzi 55, 20126 Milano, Italy \\
and INFN, sezione di Milano-Bicocca}

\emailAdd{a.hanany@imperial.ac.uk, alessandro.tomasiello@unimib.it, e.van-den-driessche24@imperial.ac.uk}

\abstract{Higgs branches of 5d $Sp(k)$ theories with $N_f$ flavours, whether at weak or strong coupling, are described by a pair of instantons transforming as pure spinors of $SO(2N_f)$. The Poisson structure is constrained by symmetry arguments and implies that these Higgs branches are algebraic integrable systems; the degeneration of the symplectic form occurs when the spinor annihilators overlap. We argue that the stratification of the Higgs branch at infinite coupling corresponds to the alignment of the instantons weights, leading to a non vanishing magnetization, and their acquisition of a mass.}
\begin{document}
\maketitle
\flushbottom

\section{Introduction}
Theories with eight supercharges occupy a sweet spot among supersymmetric ones, as they allow for a degree of mathematical control without being over-constrained. This can be seen at the level of the moduli space of vacua: with a larger number of supercharges, it is typically just flat space, whereas theories with fewer supercharges are harder to control and may not even possess a moduli space to begin with.\\
Among theories with eight supercharges, five-dimensional theories have received  less attention than those in lower dimensions three and four. Although perturbatively non-renormalizable, they were shown to admit a fixed point of the RG flow in the UV \cite{seiberg1996five} for suitable matter content. In two dimensions and below, gauge degrees of freedom are either topological or act as constraints; above six dimensions, supersymmetric theories do not admit a unitary superconformal fixed point. \\
In the study of five-dimensional supersymmetric theories, the Coulomb branch of the moduli space of vacua has received more attention than the Higgs branch, due to the popular belief that the latter is protected under RG flow. However, the chiral ring describing the latter was recently shown to be corrected at both weak and strong coupling \cite{hanany2025chiral}.\\

The aim of this paper is to consider this oft-neglected branch of vacua of five dimensional theories. For $Sp(k)$ theories with $N_f$ flavours, we will show that, on the Higgs branch at strong coupling, the only independent BPS degrees of freedom of the low-energy theory are instanton and anti-instanton operators. They both transform as pure spinors under the theory's flavour symmetry $SO(2N_f)$, and determine the spectrum at both strong and weak coupling. Moreover, we will argue that the stratification of the Higgs branch implies that the instantons, at strong coupling, acquire a mass along sub-strata of the Higgs branch where Coulomb branch directions open up.\\

The fact that instantons constitute the elementary degrees of freedom, instead of mesons or gaugino bilinears, is due to the integrable structure of the Higgs branch. We can study this integrable structure by bootstrapping the Poisson bracket using covariance under global symmetries and scaling dimensions. In particular, the Poisson bracket of the instanton and anti-instanton exhibits a peculiar non-linear structure.
Through the Liouville--Arnold theorem, we exhibit the Higgs branch as the phase space of an integrable system; the instantons parameterise a Lagrangian submanifold, and constitute the action variables in terms of which the holomorphic symplectic two-form can be written. The stratification in symplectic leaves follows from the orbits under $SO(2N_f)$ of the pure spinor pair. \\

The structure of the paper is the following: in Section \ref{sec:chiralring} we characterize Higgs branches of supersymmetric gauge theories and define more precisely the questions we aim to address. In Section \ref{sec:pure} we show that the instantons transform as pure spinors under the global symmetry and they determine the chiral ring at both strong and weak coupling. In Section \ref{sec:integrability} we identify the Higgs branch as an algebraic integrable system. In Section \ref{sec:stratification} we derive the stratification of the Higgs branch in symplectic leaves and we give a physical argument in terms of the instanton mass.

\section{5d Higgs branches}\label{sec:chiralring}
In this section we explain how Higgs branches of 5d supersymmetric gauge theories are characterized quantitatively. Recent progress in the necessary computational techniques have been described in \cite{hanany2025chiral,hanany2025higgs} and references therein.

\paragraph{Massless BPS spectrum.}
Given a Higgs branch vacuum state of a supersymmetric theory, degenerate states are obtained by applying gauge-invariant local operators, annihilated by
half of the supercharges of the supersymmetric algebra; this set is closed under product and is thus called \textit{chiral ring}. 
The vacuum expectation values of these operators parameterize the moduli space of supersymmetric vacua.
A given point on this moduli space identifies a low energy theory. When the latter is free, the moduli are non-interacting and unconstrained, the chiral ring is freely generated, and the moduli space is just flat space. The situation is much more interesting when the moduli are interacting, i.e.~there are chiral ring relations which bend and deform the moduli space of supersymmetric vacua.\\

This article focuses on the Higgs branch of theories with symplectic gauge group $Sp(k)$, with $N_f\le 2k+3$ fundamental flavours.\footnote{
$k$ denotes the rank of the group; for example, $Sp(1)$ is isomorphic to $SU(2)$. The cases $N_f=2k+4$, $2k+5$ have a symmetry enhancement phenomenon, which makes their study a bit different; we leave it to future study.} It has been studied extensively at both the weakly coupled and strongly coupled fixed points of the RG flow \cite{hanany2025chiral, hanany2025higgs}. It was found, in accordance with \cite{tachikawa2015instanton, zafrir2015instanton, Kim:2012gu}, that the independent chiral ring generators at strong coupling have the quantum numbers shown in Table \ref{tab:generators}.

\renewcommand{\arraystretch}{1.3}
\begin{table}[ht]
    \centering
    \begin{tabular}{|c|c|c|c|} \hline
       Generator  &  R-charge & $SO(2N_f)$ & $U(1)$\\ \hline \hline
        $M_{[ij]}$ & 2 & Adjoint & 0 \\ \hline
        $S$ & 2 & Singlet & 0 \\ \hline
        $I,\tilde{I}$ & $h^{\vee}$ & Spinor & $\pm 1$ \\ \hline
    \end{tabular}
    \caption{Generators of the chiral ring at infinite coupling. $h^{\vee}=k+1$ is the dual Coxeter number of $Sp(k)$. We use the convention that the R-charge is the highest weight of the $SU(2)_R$ representation.}
    \label{tab:generators}
\end{table}

At strong coupling these moduli should be regarded as elementary fields rather than composites. The generators $M_{ij}$ and $S$ are the moment maps of the global symmetries $SO(2N_f)$ and $U(1)$. The instanton $I$ and anti-instanton $\tilde{I}$ are charged under the topological $U(1)$ symmetry and the R-symmetry. Furthermore, they admit a description in the weakly coupled IR theory as local disorder operator: they create states that correspond to solutions of the instanton equation, which are localized along four dimensions and extended along one. As originally found in \cite{seiberg1996five}, their masslessness is a consequence of the infinite coupling of the theory in the UV. If we deform to finite coupling, they acquire a mass and decouple from the chiral ring. \\
The chiral ring relations at infinite coupling are shown in Table \ref{tab:infinitecouplingconstraintsGENERAL}.
\renewcommand{\arraystretch}{1.3}
\begin{table}[H]
    \centering
    \begin{tabular}{|c|c|c|c|} \hline
     Constraint  & R-charge & $U(1)$ charge \\ \hline \hline
       $ M^2_{ij}=S^2\delta_{ij}$ & 4  & 0    \\ \hline 
      $(M\gamma_{})\cdot I=SI$ & $k+3$ & +1  \\ \hline
      $\epsilon\cdot \wedge^qM=S^a \, (\wedge^{q-a}M)$ & $2q$ & 0  \\ \hline
      $ S^{h^{\vee}-c} (\wedge ^{c}M_{ij})=(I\cdot\tilde{I)}|_{\mu_{2c}}$ & $2k+2$ & 0  \\ \hline
      $\text{Sym}^2I=(I\cdot I)|_{\mu_{N_f}^2}$ & $2k+2$ & +2  \\ \hline
    \end{tabular}
    \caption{Chiral ring constraints for $Sp(k)$ with $0<N_f \leq 2k+3$ flavours in the vector representation, in the infinite coupling limit. The parameters $a$ and $q$ are related by $N_f=2q-a$, with $q \leq k+1$; furthermore $c\leq h^{\vee}$. The second and last constraint appear also for $\tilde{I}$ but are here omitted for readability. The dot product represents an appropriate contraction of the vector or spinor indices.}
\label{tab:infinitecouplingconstraintsGENERAL}
\end{table}
Although natural from a representation-theoretic point of view, these chiral ring relations lack a physical explanation. It is unclear why they arise or what they imply, whether they are all independent or secretly related to one another. Furthermore, only the relations up to R-charge $2k+2$ were explored, in a perturbative expansion, however it is unclear whether independent relations arise at higher order too. The question:
\begin{itemize}
    \item \textbf{Question 1.}
What is the origin of the observed chiral ring relations?
\end{itemize}
is one of the two issues that this paper aims to address.\\

A further insight into the structure of the chiral ring is offered by the construction of the Higgs branch as moduli space of dressed instanton operators, as introduced in \cite{hanany2025higgs}. Indeed, the elements of the chiral ring can be counted through a generating function \cite{hanany2014highest}, which is known as \textit{highest weight generating function} (HWG) since, among all BPS operators, it only generates their highest weights under global symmetries. When the desired 5d Higgs branch can be generated as the moduli space of dressed instanton operators, the HWG can be written as product of an instanton contribution and a dressing factor. The instanton contribution is the sum over all instanton topological charges, while the dressing factor is the HWG of the Higgs branch at weak coupling.\\

The advantage of this formulation is in revealing that the holomorphic functions on the Higgs branch behave as a union of two cones. Indeed, the HWG for the theories we are studying can be written, schematically, as:
\begin{equation}\label{eq:dressedinstantons}
    \text{HWG}=\bigg(\text{PE}[I]+\text{PE}[\tilde{I}]-1\bigg)\cdot \text{PE}[M]\cdot \text{PE}[S] \ ,
\end{equation}
where PE stands for the plethystic exponential, while $I,\tilde{I},M$ and $S$ represent the chiral ring generators; for details, we refer to \cite{hanany2025higgs}. Eq.~\ref{eq:dressedinstantons} just means that the chiral ring can be divided into topological sectors according to the instanton charge. There are sectors of positive instanton charge, spanned by $\text{PE}[I]$, and negative instanton charge $\text{PE}[\tilde{I}]$, with the $-1$ in Eq.~\ref{eq:dressedinstantons} representing the identity, present in both types of sectors. Each topological sector of charge $q$ is populated by operators which are the product of three contributions: an instanton of charge $q$, transforming in a representation of the global symmetries, the chiral ring of the weak coupling theory, represented by $\text{PE}[M]$, and powers $\text{PE}[S]$ of the gaugino bilinear.\\

The chiral ring generators in Tab.~\ref{tab:generators} are a redundant set of coordinates on the Higgs branch, in the sense that their number is greater than the dimension of the Higgs branch. 
The insight given by Eq.~\ref{eq:dressedinstantons} is that we should think of the 5d Higgs branches in question as the union of two orthogonal cones, \textit{padded} with mesons and gaugino bilinear $S$ in such a way to lose orthogonality. Indeed, from the fourth constraint in Tab.~\ref{tab:infinitecouplingconstraintsGENERAL}, the product of instantons and anti-instantons is nonzero, hence the two cones cannot be orthogonal. 

\paragraph{Stratification.}
Higgs branches of theories with eight supercharges are \textit{symplectic singularities} \cite{Beauville:1999jhe,Namikawa2001,Kaledin2006}: they are singular varieties $X$ with a hyperK\"ahler structure on their smooth part, such that its pull-back to any resolution $\tilde X\to X$ can be extended over all of $\tilde X$. A hyperK\"ahler structure is given by three complex structures $J_{1,2,3}$ obeying the quaternionic algebra:
\begin{equation}
    J_i J_j=-\delta_{ij}\mathbf{1}+\epsilon_{ijk}J_k\,.
\end{equation}
They correspond to weights of an $SU(2)$ adjoint representation, corresponding to the R-charge $SU(2)_R$ in 5d. A choice of a complex structure corresponds to a choice of complex coordinates, in terms of which we define holomorphic functions such as the moduli.\\
One can also define a \emph{holomorphic symplectic} form $\omega \equiv (J_1+ \ii J_2).g$, where $g$ is the metric. It is holomorphic in the sense that it is $(2,0)$ with respect to $J_3$, and symplectic in that it is closed, and non-degenerate: $\wedge^{d/4}\omega$ (with $d$ the real dimension) is a holomorphic volume form. Its inverse $\omega^{-1}$ is a Poisson structure, defining a Poisson bracket:
\begin{equation}
    \{f,g\}=(\omega^{-1})^{ij}\partial_i f \partial_j g \ ,
    \label{eq:generalpoisson}
\end{equation}
where $f,g$ are holomorphic functions with respect to $J_3$. This $\omega^{-1}$ can be extended to a Poisson structure on all of $X$. As in the real case, 
$X$ can now be canonically foliated into \textit{symplectic leaves}: submanifolds on which the restriction of $\omega^{-1}$ is non-degenerate (so that they are symplectic), and such that their dimension equals the rank of the Poisson tensor at any of their points.
Any two points on a symplectic leaf can be connected by an integral curve of any Hamiltonian vector field $X_f$:
\[
X_f \equiv \text{d}f \lrcorner \,\,\omega^{-1} \ ,
\]
with $f$ any holomorphic function. It follows in particular that no point outside the symplectic leaf can be reached through the flow of Hamiltonian vector fields on the leaf itself.
\\

Allowing the Poisson bracket to degenerate has important physical consequences. In particular, for theories with Lagrangian description, the pattern of degeneration of $\omega^{-1}$ determines the pattern of Higgs breaking of the gauge symmetry, as detailed in \cite{bourget2020higgs}. This means that symplectic leaves are described by the residual gauge symmetries after the Higgs mechanism. In theories that do not admit a Lagrangian description, symplectic leaves still describe different phases of the theory, namely different sets of interacting massless moduli.\\

The different phases of the theory can be neatly encoded in the so-called \textit{Hasse diagram}, which visually describes the stratification of the Higgs branch. The highest point in a Hasse diagram corresponds to the largest leaf of the theory, which is the locus of points where the gauge group is maximally broken, in theories with a Lagrangian formulation. By tuning some moduli to zero, we restrict ourselves to a smaller locus identified by a further degeneration of $\omega^{-1}$, i.e.~to a smaller leaf. After tuning more and more moduli to zero, we reach the origin of the Higgs branch. We thus see that the stratification is a matryoshka-like structure of nested singularities, in that it admits a partial ordering with the closure of any leaf $\mathcal{O}$ including all smaller ones:
\[
\bar{\mathcal{O}}\supseteq \mathcal{O}_1 \supset \mathcal{O}_2 \supset \dots \mathbf{1} \ .
\]
with $\mathbf{1}$ representing the origin of the space.\\

If a Lagrangian formulation is available, the Hasse diagram can be computed entirely from the Higgsing of the gauge group. More generally, a conjectural algorithm \cite{cabrera2018quiver} has been proposed to draw Hasse diagrams, called \textit{quiver subtraction}. This algorithm provides a partial ordering of Coulomb branches of $3d$ $\mathcal{N}=4$ theories, through a brane construction known as the \emph{Kraft--Procesi transition}. The algorithm claims that this partial ordering corresponds to the partial ordering of symplectic leaves. While this claim has been tested on Lagrangian theories, and has led to a wealth of extensions in recent years (see for example \cite{bennett2026quotient, Bennett:2024loi}) it remains hypothetical when non Lagrangian theories come into play, such as 5d strongly coupled Higgs branches. The question:
\begin{itemize}
    \item \textbf{Question 2.}
Can we independently derive the Higgs branch stratification?
\end{itemize}
which is the second issue this project addresses, aims to substantiate the claims of quiver subtraction.\\

It might seem that the two questions we have posed are independent from each other. We will instead derive the Higgs branch stratification precisely from the chiral ring generators and relations; it is generally unclear how to carry out this step. Moreover, we will see that the projector on the two instanton cones plays a crucial role in determining the stratification.\\
Furthermore, although the massless BPS spectrum and the stratification already provide an abundant description of the Higgs branch, the story does not end here, and a further characterization might involve computing the correlation functions of the operators in the theory. In this context, we point out \cite{beem2017deformation, beem2020secondary} that the OPE of two chiral ring elements always includes a nonsingular term, given by the chiral ring relations, but if we consider deformation quantization of the chiral ring, singular terms are present as well. These terms are singular, as they depend on the distance between the insertions points; the first subleading term is proportional to the Poisson bracket of the chiral ring elements in question. We thus see that determining the Poisson bracket of the chiral ring is not only interesting for the stratification of the Higgs branch, but for the dynamics of the theory as a whole.

\section{Pure spinors}
\label{sec:pure}

In this section, we will focus on the interpretation of instantons as pure spinors. We will show that the instantons are generically the only independent degrees of freedom; we will also see how all chiral ring relations are secretly a consequence of the fourth constraint in Tab.~\ref{tab:infinitecouplingconstraintsGENERAL}.

\subsection{Definitions}\label{sub:pure}

We will mostly use an index-free notation for spinors. The transpose of a spinor needs to be combined with an intertwiner matrix: given $B$ such that $B\gamma_i^t = \pm B \gamma_i $, one can define $\eta^{\rm c} \equiv\eta^t B$; the inner product
\begin{equation}\label{eq:inner}
    \eta_1 \cdot \eta_2 \equiv \eta_1^{\rm c} \eta_2 
\end{equation}
is now invariant under $SO(d=2N_f)$. We also define $|\eta|^2 \equiv \eta \cdot \eta$. In terms of indices, the matrix $B$ can be written as $\delta_{\alpha\beta}$ when the spinorial representation is real (when $N_f=4M$, $M\in \mathbb{Z}$); 
$\Omega_{[\alpha\beta]}$ when it is pseudoreal ($N_f=4M+2$); $\delta_{\alpha}^{\beta}$ when it is complex ($N_f$ odd). In the first two cases $\eta^{\rm c}$ has the same chirality as $\eta$; in the last, the opposite. All in all, $\eta_1 \cdot \eta_2 = \sigma \eta_2 \cdot \eta_1$, where $\sigma$ is $-1$ when $N_f=4M+2$, and $+1$ otherwise. \\

As we mentioned, the last relation in Tab.~\ref{tab:infinitecouplingconstraintsGENERAL} defines a \textit{pure spinor} (see for example \cite[Sec.~2.4]{tomasiello2022geometry}). Indeed:
\begin{equation}\label{eq:Ipurespinor}
    \text{Sym}^2I=(I\cdot I)|_{\mu_{N_f}^2} \ ,
\end{equation}
is a possible definition of pure spinor. Indeed Eq. ~\ref{eq:Ipurespinor} is the representation theoretic translation of the condition:
\begin{equation}\label{eq:pure-bil}
	I \cdot  \gamma_{i_1\ldots i_q}  I = 0  \qquad  \forall q \neq N_f \,.
\end{equation}
Here and in what follows, indices from the middle of the Latin alphabet (such as $i$, $j$) range from 1 to $d=2N_f$. 

An alternative definition involves the annihilator of a spinor, defined as:
\begin{equation}\label{eq:ann}
	\mathrm{Ann}(I)\equiv \left\{ v \in \mathbb{C}^d \ |\  v^i \gamma_i I = 0 \right\}\,.
\end{equation}
A pure $I$ now satisfies the condition
\begin{equation}\label{eq:pure}
	\mathrm{dim}_{\mathbb{C}}\mathrm{Ann}(I) = d/2= N_f\,.
\end{equation}
Another way of writing Eq.~\ref{eq:ann} is
\begin{equation}\label{eq:Pi-I}
	\Pi_i{}^j \gamma_j I = 0 \,,
\end{equation}
where $\Pi$ is a projector on $\mathrm{Ann}(I)$. This is often expressed as $\Pi=\frac12(1- \ii J)$, with $J$ a complex structure. Still another possibility is to introduce complex coordinates $z^a$ in the $2N_f$-dimensional Euclidean space, with $a=1,\ldots,N_f$ a holomorphic index. Eq.~\ref{eq:Pi-I} then reads $\gamma_a I=0$. 

We will show below that the definition Eq.~\ref{eq:pure-bil} and the condition Eq.~\ref{eq:pure} are equivalent.

\paragraph{Orbit.} Up to rescaling, all pure spinors of a given chirality live in the same orbit: they can be obtained from one another by a rotation. In other words, pure spinor orbits are characterized by norm and chirality. 

It is useful to work in the oscillator basis (see for example \cite[Sec.~2.2.1]{tomasiello2022geometry}), where gamma matrices read:
\begin{equation}\label{eq:basiz}
\begin{split}
	\gamma_{z^1} &= \sigma_z\otimes 1 \otimes 1 \otimes \ldots \otimes 1  \, ,\\
	\gamma_{z^2} &= \sigma_3\otimes \sigma_z \otimes 1 \otimes \ldots \otimes 1 \,, \\ 
	\vdots&\\
	\gamma_{z^{d/2}} &= \sigma_3 \otimes \sigma_3 \otimes \sigma_3 \otimes \ldots \otimes \sigma_z\,,
\end{split}
\end{equation}
where $\sigma_z= \frac12(\sigma^1 - \ii \sigma^2)$. A basis for spinors can now be written as 
\begin{equation} \label{eq:|a>}
    | s_1 \ldots s_{d/2=N_f} \rangle 
\end{equation}
where $s_1\in \{+,-\}$, with the understanding that $\gamma_{\bar z} |+ \rangle =0 = \gamma_z |- \rangle$. In general, we use a short-hand notation to denote sign repetitions: for example, 
$\underset{p}{\underbrace{+\, \dots  \, +}}\equiv +^p$. 
We can take the conjugation matrix $B=\sigma_1\otimes \sigma_2 \otimes \sigma_1\otimes\ldots$, which reverses all signs. So for example for $\eta= |+^{N_f}\rangle$, $\eta'=|-^{N_f}\rangle$  the inner product reads $\eta'\cdot \eta = (\eta')^t B \eta= \langle+^{N_f}|+^{N_f}\rangle = \ii^{\lfloor N_f/2\rfloor}$.

Let us now see why Eq.~\ref{eq:pure} implies Eq.~\ref{eq:pure-bil}. We take the $SO(2N_f)$ rotation that maps the $d/2=N_f$ elements of the annihilator to the vector fields $\partial_{\bar z^{\bar a}}$. Now $\gamma_{\bar z^{\bar a}}I=0$, so $I$ must be proportional to
\begin{equation}
    |\underset{N_f}{\underbrace{+\, \dots  \, +}\rangle}= | +^{N_f} \rangle\,.
\end{equation}
Incidentally, this shows that all pure spinors belong to the same orbit. Now the bilinears Eq.~\ref{eq:pure-bil} read $\langle -^{N_f} | \gamma_{i_1\ldots i_q} |+^{N_f}\rangle$. We see that $\gamma_{i_1\ldots i_q}$ has to contain exactly $N_f$ annihilators to turn $|+^{N_f}\rangle$ into $|-^{N_f}\rangle$; any other choice leads to a zero result. This is exactly Eq.~\ref{eq:pure-bil}.

In this basis, it is also easily seen that the stabilizer of a pure spinor is $SU(d/2=N_f)$ (see for example \cite[Sec.~2.4.4]{tomasiello2022geometry}). Adding the complex rescaling, we find that the space of pure spinors is
\begin{equation}\label{eq:coset}
    Q=\mathbb{C}^*\times\dfrac{SO(2N_f)}{U(N_f)}\,, \qquad
\mathrm{dim}_{\mathbb{C}} Q =  \dfrac{N_f (N_f-1)}{2}+1\,.
\end{equation}

\subsection{Pure spinor pairs}\label{sub:purespinorpair}

\paragraph{Bispinors and forms.}

A \textit{bispinor} is a map from the space of spinors to itself. A particularly important class of bispinors (which constitutes a basis) is given by those of rank one: 
\begin{equation}\label{eq:Phi}
    \Phi \equiv I \otimes \tilde I^{\rm c}\,,
\end{equation}
where $I$, $\tilde I$ are two spinors and ${}^{\rm c}$ was defined above Eq.~\ref{eq:inner}. This acts as $I \otimes \tilde I^{\rm c}: I' \mapsto I (\tilde I \cdot I') $.\footnote{This is like a Hilbert space projector $| \psi \rangle \langle \tilde\psi|: |\psi' \rangle \mapsto | \psi \rangle  
\langle \tilde\psi| \psi' \rangle$.} Another notable basis for bispinors is given by the $\gamma_{i_1\ldots i_q} $, $q=0,\ldots,d$. Projecting on this basis gives the \emph{Fierz identity}:
\begin{equation}\label{eq:fierz}
    I \otimes \tilde I^{\rm c} = \sum_{q=0}^d\frac1{2^d q!} (\tilde I \cdot \gamma_{i_q\ldots i_1} I) \gamma^{i_1\ldots i_q}\,.
\end{equation}

A bispinor can be mapped to a polyform by inverting the Clifford map
\begin{equation}\label{eq:clifford}
    \dd x^{i_1}\wedge \ldots \wedge \dd x^{i_k}\ \mapsto\ \gamma^{i_1\ldots i_k}\,.
\end{equation}
With an abuse of language, we will often confuse a polyform with the corresponding bispinor and viceversa.
In this language, Eq.~\ref{eq:pure-bil} can be expressed as saying that $I \otimes I^{\rm c}$ is a $N_f=d/2$-form (a \emph{holomorphic volume form}).

 A polyform (and hence a bispinor) can be considered a spinor for the \textit{doubled Clifford algebra} $\mathrm{Cl}(d,d)$, whose gamma matrices are simply $\iota_i\equiv \iota_{\partial_i} $ and $\dd x^i \wedge$. There is a linear map between the latter and the $\mathrm{Cl}(d)$ gamma matrices acting from the left and from the right:
\begin{equation}\label{eq:gG}
	\stackrel\to {\gamma^i} = \dd x^i \wedge + g^{ij}\iota_j \, ,\qquad
	\stackrel\leftarrow {\gamma^i} = (\dd x^i \wedge - g^{ij}\iota_j)(-1)^\mathrm{deg} \,,
\end{equation}
where deg is the degree of the form we are acting on. For example, the first in Eq.~\ref{eq:gG} implies $\gamma_i \gamma^{i_1\ldots i_q}= \gamma_i^{i_1\ldots i_q} + q\, \delta_i^{[i_1} \gamma^{i_2\ldots i_q]}$.

Let us now show that Eq.~\ref{eq:pure-bil} implies Eq.~\ref{eq:pure}. Consider the case where $\tilde I=I$, i.e.~$\Phi= I \otimes I^{\rm c}$, which is an $N_f$-form by Eq.~\ref{eq:fierz}.  Given an $(N_f-1)$-form $\xi$, define the contraction operator $\iota_\xi \equiv \frac1{(N_f-1)!}\xi^{i_1\ldots i_{N_f-1}}\iota_{i_1}\ldots \iota_{i_{N_f-1}}$. Using Eq.~\ref{eq:gG} and Eq.~\ref{eq:pure-bil}, one finds $(\iota_\xi \Phi) \Phi=0$, because all terms have fewer than $N_f-1$ gamma matrices between $I^{\rm c}$ and $I$. Now $\iota_\xi \Phi$ is a one-form, and we can use Eq.~\ref{eq:gG} again to see that $(\iota_\xi \Phi) \Phi$ is a sum of a $(N_f+1)$- and a $(N_f-1)$-form, which have to vanish separately; the latter is $\iota_\xi \Phi \wedge \Phi=0$. These are the Pl\"ucker relations: they imply that $\Phi$ is decomposable, $\Phi= h_1 \wedge \ldots \wedge h_{N_f}$. Now $0=h_a \wedge \Phi$, which using the definition of $\Phi$ gives the desired $N_f$-dimensional ${\rm Ann}(\Phi)$. 

\paragraph{Type.} Take now $I$, $\tilde I$ to be pure. The discussion surrounding Eq.~\ref{eq:gG} implies that $\mathrm{Ann}(I \otimes \tilde I^{\rm c})$ has dimension $d/2+d/2=d$; in turn, $I \otimes \tilde I^{\rm c}$ is itself a pure spinor for $\mathrm{Cl}(d,d)$. As a form, it can be written as\footnote{Write $\mathrm{Ann}(\Phi)= \{ M_i{}^j \iota_j- N_{ij} \dd x^j \wedge\}$. If $M$ has rank $d-t$, by linear combinations over the first index we can arrange for it to be block diagonal with blocks $1_{d-t}$, $0_{t}$. Then the $ h_i\equiv N_{ij}\dd x^j$ annihilate $\Phi$ for $i>t$; define  $\alpha_t \equiv h_{d-t+1} \wedge \ldots \wedge h_d$. Also, $\iota_i - N_{ij}\dd x^j$ annihilate $\Phi$ for $i<1$. Overall this fixes Eq.~\ref{eq:Phi-typet}, with $\iota_v N=-\iota_v \beta$ for $v$ such that $\iota_v \alpha_t=0$.}
\begin{equation}\label{eq:Phi-typet}
	\Phi= I \otimes \tilde I^{\rm c} = \alpha_t \wedge \ee^{ \beta_2}\,
\end{equation}
where $\alpha_t$ is a \emph{decomposable} $t$-form (it can be written as a wedge product of one-forms) and $\beta_2$ is a two-form respectively. We can define the \emph{type}: $t=t(\Phi)=t(I,\tilde I)$, which is a function of $I$ and $\tilde{I}$: it is the lowest form rank appearing in $\Phi$. 

To gain some intuition, let us consider some examples. As we mentioned, we can take $I=|+^{N_f}\rangle$ without loss of generality.  For spinors in the basis Eq.~\ref{eq:|a>}, the type is the number of $+$ signs:
\begin{equation}\label{eq:t+^N}
t \left(|+^{N_f}\rangle,| s_1 \ldots s_{N_f} \rangle\right)= \#\{s_q = +\}\,.
\end{equation}
We will sometimes say that the two spinors have an \emph{overlap} $s_q$.
For example, $\tilde I = |-^{N_f} \rangle$ gives $t=0$; if $\tilde I = I =|+^{N_f} \rangle$ we see that $t=N_f$, consistent with Eq.~\ref{eq:pure-bil}. More generally, if $\tilde I$ is a linear combination of spinors in the basis Eq.~\ref{eq:|a>}, $t$ is the minimum number of $+$ signs in any component.\footnote{To see this, take $\alpha_t = h_{d-t+1} \wedge \ldots \wedge h_d$ as before; there are $t$ annihilators of the form $h_i\wedge$. Inverting Eq.~\ref{eq:gG}, this means $h_i \gamma^i \Phi = 0 = \Phi h_i \gamma^i$, which implies $h_i \gamma^i I = h_i \gamma^i \tilde I=0$. So there are $t$ linear combinations of gamma matrices that annihilate both $I$ and $\tilde I$.} All this implies in particular that\footnote{For a pure $\mathrm{Cl}(d,d)$ spinor that is not defined via Eq.~\ref{eq:Phi}, it is possible to achieve higher rank: for example, $\mathrm{vol}_d$ is pure ($\mathrm{Ann}(\mathrm{vol}_d)= {\rm Span}\{ \iota_i\}= T^*$) and has $t=d$.}
\begin{equation}\label{eq:td2}
    t\le d/2=N_f\,.
\end{equation}

\paragraph{Space of pairs.}

We saw earlier that the space of pure spinors with a given norm and chirality is a single $SO(d)$ orbit. If we consider instead the space of pairs $(I,\tilde I)$, our earlier discussion shows that the situation is much more complicated. Indeed, rotations do not change $t$, so pairs $(I,\tilde I)$ with different $t$ belong to different $SO(d)$ orbits. A higher value of $t$ is less and less generic, since it corresponds to tuning to zero more and more forms in Eq.~\ref{eq:Phi-typet}. In particular, the generic case is type $t=0$ since it corresponds to the largest orbit of pairs $(I,\tilde I)$. The relation between type and dimension of the orbit is made clear in Sec.~\ref{sec:stratification}.

In general, the type is not the only $SO(d)$ invariant. As an example, in $d=2N_f=6$, consider $I=|{+++} \rangle$, $\tilde I = \cos \theta |{---} \rangle + \sin \theta |{++-} \rangle$. This pair is of type 0, but $\frac{\tilde I \cdot I}{|I||\tilde I|}=\cos\theta$ is also an invariant.

\paragraph{Hodge duality.}

Let $\gamma= c \gamma_1\ldots \gamma_{d=2N_f}$; the constant $c$ is usually chosen so that $\gamma$ is hermitian and squares to one. Then we have the Hodge star relation
\begin{equation}\label{eq:gamma*slash}
	\gamma \alpha = c * \lambda \alpha\,.
\end{equation}
Here $\lambda \alpha_k \equiv (-1)^{\lfloor \frac k2 \rfloor} \alpha_k = (-1)^{\frac{k(k-1)}2} \alpha_k $.

For us $I$ has positive chirality; so we have the self-duality property
\begin{equation}\label{eq:*Phi}
	\Phi = c* \lambda \Phi \,.
\end{equation}
In other words, $*\Phi_k = c^{-1} (-)^{\lfloor \frac k2 \rfloor} \Phi_k$. This self-duality property also implies again the bound Eq.~\ref{eq:td2}.

\subsection{Chiral ring origin}
\label{sub:ring}

We will now see that only some orbit types are allowed by the chiral ring relations. 

\paragraph{Zero coupling.} As a warm-up, consider the zero coupling case, where only mesons are present. The chiral ring relations read here
\begin{equation}\label{eq:w-rel}
    M^2=0 \,,\qquad \wedge^{k+1}M=0\,.
\end{equation}
Recall that $M$ is a $2N_f$-dimensional antisymmetric matrix.
Here we can use the classification of nilpotent orbits of $SO(2N_f)$. Recall that these orbits are characterized by the decomposition into Jordan blocks; the list of their sizes gives a partition of $d=2N_f$, such that even numbers appear an even number of times. Since the nilpotency order is two, the size of the blocks is $\leq 2$. Such a partition can be denoted as
\begin{equation}\label{eq:M-orbits}
    [2^{2i},1^{2(N_f-2i)}]\,.
\end{equation}
In this orbit, ${\rm rank}M=2i$; by Eq.~\ref{eq:w-rel}, the rank satisfies ${\rm rank}M\leq {\rm min}(2k,N_f)$. If $k\ge N_f/2$ and $N_f$ is even, we can reach $2i=N_f$, in which case there are actually two orbits:
\begin{equation}\label{eq:2^d/2}
	[2^{N_f}]_+ \, ,\qquad [2^{N_f}]_-\,.
\end{equation}

Let us give some examples to understand these orbits more concretely. A $4\times 4$ matrix representing $r=2$ blocks of size 2 can be brought to the form $m_4 = \dd z^1 \wedge \dd z^2$, with $z^1=x^1+\ii x^3$, $z^2=x^2 + \ii x^4$ complex coordinates. For ${\rm rank} M = 2i$, we have
\begin{equation}\label{eq:Mz}
	M = \dd z^1 \wedge \dd z^2 + \ldots + \dd z^{2i-1} \wedge \dd z^{2i}\,.
\end{equation}
Different choices of complex coordinates lead to superficially different $M$. For $2i<N_f$, these are actually in the same orbit. For example, one can map $M'=\dd z^1 \wedge \dd z^2 + \ldots + \dd z^{2i-1} \wedge \dd \bar z^{2i}$ (with a single barred coordinate) to Eq.~\ref{eq:Mz} by the element of $SO(2N_f)$ that reverses sign to ${\rm Im} z^{2i}$ and to one of the coordinates not contained in $z^1,\,\ldots,z^{2i}$. In the case $2i=N_f$, all coordinates are contained in $z^1,\,\ldots,z^{N_f}$; so mapping $M'=\dd z^1 \wedge \dd z^2 + \ldots + \dd z^{N_f-1} \wedge \dd \bar z^{N_f}$ to Eq.~\ref{eq:Mz} is impossible. This is the case more generally for any $M'$ obtained from Eq.~\ref{eq:Mz} by barring an odd number of $z^a$. This leads to Eq.~\ref{eq:2^d/2}.

\paragraph{Infinite coupling: relations.}
We now move on to infinite coupling. The relations in Tab.~\ref{tab:infinitecouplingconstraintsGENERAL} read:
\begin{subequations}\label{eq:rel} \begin{align} \label{eq:M2} &M^2_{ij} = S^2 \delta_{ij} \,;\\ \label{eq:MI} &M_{ij}\gamma^j I = S \gamma_i I \, ,\qquad M_{ij}\gamma^j \tilde I= S \gamma_i \tilde I \,;\\ \label{eq:*M} &* \wedge^q M = S^{2q-N_f} \wedge^{N_f-q} M\,; \qquad (q\leq k+1)\\ \label{eq:II} &\tilde I \cdot \gamma_{(2q)} I = S^{k+1-q} \wedge^q M \,; \quad\qquad (q\leq k+1)\\ \label{eq:pure-rel} & I \cdot \gamma_{(2q)} I = \tilde I \cdot \gamma_{(2q)} \tilde I = 0\,,\qquad \qquad\forall q \neq N_f\,, \end{align} \end{subequations}
up to numerical coefficients, to be determined in the following. $I$ has chirality $+1$, and $\tilde I$ has chirality $(-1)^{N_f}$. By Eq.~\ref{eq:fierz}, $\Phi$ is always an even form. 
For example, Eq.~\ref{eq:II} is the $2q$-form part of Eq.~\ref{eq:Phi-typet}, with  $\gamma_{(2q)}\equiv \frac1{(2q)!}\gamma_{i_1\ldots i_{2q}} \dd z^{i_1} \wedge \ldots \wedge \dd z^{i_{2q}}$. The bound $q\le k+1$ in Eq.~\ref{eq:*M} and Eq.~\ref{eq:II} is a result of exploring the Hilbert series perturbatively up to R-charge $2k+2$ in \cite{hanany2025chiral}. We will argue shortly that this bound can in fact be dropped.

Eq.~\ref{eq:pure-bil}, Eq.~\ref{eq:pure} are equivalent to $I$ and $\tilde I$ being pure spinors. We will now interpret the rest of the relations, dividing the analysis into subcases $S\neq 0$ and $S=0$.

\paragraph{Orbits with $S\neq 0$.}
$M$ is no longer nilpotent. By Eq.~\ref{eq:fierz}, the left-hand sides of Eq.~\ref{eq:II} are proportional to the $2q$-form parts of $\Phi$. The $q=0$ part of Eq.~\ref{eq:II} reads $\tilde I\cdot I = S^{k+1}$ (possibly after rescaling $S$). For $S\neq 0$ we conclude  $\tilde I\cdot I\neq 0$. This is the zero-form part of $\Phi= I \otimes \tilde I^{\rm c}$, so the type is
\begin{equation}
    t(I, \tilde I)=0\,.
\end{equation}
By the discussion in the previous subsection, this is the generic case, namely corresponding to the bispinor orbit of highest dimension. For $q>0$, Eq.~\ref{eq:II} are proportional to $\wedge^q M$. In view of Eq.~\ref{eq:Phi-typet}, we get
\begin{equation}\label{eq:Phi-Sneq0}
	\Phi =  S^{k+1} \ee^{M/S}\,.
\end{equation}
This fixes all the proportionality factors in Eq.~\ref{eq:II}. We see, moreover, that the bound $q\le k+1$ can be dropped.
 
By construction, the annihilator of the bispinor is $\mathrm{Ann}(\Phi)= {\rm Span}\{ \iota_i - M_{ij}/S \dd x^j \wedge\}$; using the map between $\{\iota_i,\dd x^i\}$ and the gamma matrices in Eq.~\ref{eq:gG}, we obtain $(1+ M/S)_{ij} \gamma^j I =0 = (1- M/S)_{ij} \gamma^j \tilde I$. These are just Eq.~\ref{eq:Pi-I} of the chiral ring, whose proportionality constants we have now fixed. Next, we define $\Pi_{ij}=S \delta_{ij}+M_{ij}$, so that $\Pi_{ij} \gamma^j I=0$; it follows that $0=\{\Pi_{ik}\gamma^k,\Pi_{jl}\gamma^l\} I =2\Pi_{ik}\Pi_{jl}\delta^{kl}I= 2\Pi_i{}^k\Pi_{jk}I$, and then $0=\Pi \,\Pi^t=S^2 1 - M^2$, which is Eq.~\ref{eq:M2}. Finally, Eq.~\ref{eq:*M} is expressing the Hodge duality of the bispinor, namely Eq.~\ref{eq:*Phi}. In conclusion, for $S\neq 0$ the only non-redundant relations are Eq.~\ref{eq:II} for $q=0,2$, and the condition on the purity of the spinors Eq.~\ref{eq:pure-rel}: all other chiral ring relations are implied.
 
\paragraph{Orbits with $S=0$ and $I$, $\tilde I\neq 0$.}\label{par:S=0}

Eq.~\ref{eq:M2} again gives us that $M$ is nilpotent of order two. Eq.~\ref{eq:MI} becomes $M_{ij}\gamma^j I=M_{ij}\gamma^j \tilde I=0$. It is again of the form of the annihilator of the instanton, as in Eq.~\ref{eq:Pi-I}; now $M$ is directly proportional to $\Pi_I$ and $\Pi_{\tilde I}$. If we identify the complex coordinates in Eq.~\ref{eq:Pi-I} with those in Eq.~\ref{eq:Mz}, this means $\gamma^a I = \gamma^a \tilde I=0$, with $a$ a holomorphic index now ranging only from 1 to $r={\rm rank}M$ (which is even). So an $M$ of the form Eq.~\ref{eq:Mz} would fix $I$ and $\tilde I$ to have the first $r$ signs to be $+$, or in other words they belong to the linear combination of spinors whose first $r$ signs are $+$:
\begin{equation}\label{eq:IIt-pluses}
	I, \tilde I \in {\rm Span}\Big\{| +^{r} s_{r+1}\ldots s_{d/2} \rangle\Big\}\,.
\end{equation}
Conjugating some of the $z^a$ in Eq.~\ref{eq:Mz} would flip the corresponding signs in Eq.~\ref{eq:IIt-pluses}. In particular, by Eq.~\ref{eq:t+^N}, the type of the bispinor satisfies $t(I,\tilde I)\ge {\rm rank}M$.

Let us now consider the spinor bilinears in Eq.~\ref{eq:II}. For $q\le k+1$, considering $S=0$ sets to zero the $2q$-form part of $\Phi$, giving
\begin{equation}\label{eq:tS=0}
    2(k+1) \le t(I,\tilde I) \le N_f\,,
\end{equation}
where we also recalled Eq.~\ref{eq:td2}. The $q=k+1$ part of Eq.~\ref{eq:II} reads $\tilde I\cdot \gamma_{(2k+2)} I = \wedge^{k+1} M$. We saw below Eq.~\ref{eq:Phi-Sneq0} that these relations can be extended to the case $q>k+1$ as well, where they read $\wedge^{k+2} M= 0$: in particular, setting $S=0$ implies ${\rm rank} M \le 2(k+1)$.

\begin{itemize}
	\item If $\mathrm{rank}(M)= 2(k+1)$, then $ \wedge^{k+1} M \neq 0$. Hence, the bispinor has to be of the form
\begin{equation}
	\Phi= \wedge^{k+1} M  \wedge \ee^{\beta_2}
\end{equation}
for some two-form $\beta_2$; so  $t=2(k+1)$. The value of  $\beta_2$  is inconsequential in what follows. Eq.~\ref{eq:*M} is again redundant.

    \item If $\mathrm{rank}(M)\le 2k$, then $\wedge^{k+1}M=0$. Now $t\ge 2(k+2)$: $\Phi$ is of the form Eq.~\ref{eq:Phi-typet}, with $\alpha_t$ now no longer necessarily related to $M$. This case is not relevant for our applications, since it would require $N_f\ge 2k+4$, and our analysis in this paper is restricted to $N_f\le 2k+3$.

\end{itemize}

\paragraph{Orbits with $S=0$, and $\tilde I=0$ or $I=0$.}
\label{par:I=0}

We now consider separately the case where either $I=0$ or $\tilde I=0$:
here $\Phi=0$, the concept of type does not really apply. While $M$ is still nilpotent of order two, the $q=k+1$ part of Eq.~\ref{eq:II} reads $ \wedge^{k+1} M = 0$, hence implying that ${\rm rank}M\le 2k$. 

\begin{itemize}

\item If $\mathrm{rank}(M)< N_f$, there is no obstruction to have either $I\neq 0$ or $\tilde I \neq 0$; it is again of the form Eq.~\ref{eq:IIt-pluses}.

\item If $\mathrm{rank}(M)= N_f$ (and thus only if $N_f\le 2k$), a possible problem is encountered. We saw in Eq.~\ref{eq:2^d/2} that when $N_f$ is even the nilpotent orbit of $M$ splits in two chiralities ($[2^{N_f}]_-$ and $[2^{N_f}]_+$).  
\begin{itemize}
    \item If $N_f$ is odd, then $I$, $\tilde I$ have opposite chirality; one of them can always exist. 
    \item If $N_f$ is even and $M\in [2^{N_f}]_+$: a possible $M$ is Eq.~\ref{eq:Mz} with $2i=N_f$. $I$ and $\tilde I$ both have positive chirality; again one of them can exist.
    \item If $N_f$ is even and $M\in [2^{N_f}]_-$, we have $I=\tilde I=0$. Here $M$ needs to have an odd number of barred coordinates: for example $M=\dd z^1 \wedge \dd z^2 + \ldots + \dd z^{N_f-3} \wedge \dd z^{N_f-2}+\dd z^{N_f-1} \wedge \dd \bar z^{N_f-2}$, with a single bar. This would require $I$ and $\tilde I$ to have negative chirality. But $I$ and $\tilde I$ are both of positive chirality, so they cannot exist. In this branch, only $M\neq 0$.    
\end{itemize} 

\end{itemize}

The branching we just observed for $N_f\le 2k$ explains the structure seen for example in \cite{bourget2020magnetic} for $k=2$, $N_f=4$, with two cones of quaternionic dimension 6 and 7. The former is $\{M\in [2^2]_-,I=\tilde I=0\}$. The latter contains the orbits with non-zero spinors; the generic case is $S\neq 0$, with $t=0$. Counting the dimensions of $I$, $\tilde I\neq 0$ with Eq.~\ref{eq:coset} we find indeed ${\rm dim}_{\mathbb{H}}= 7$. We will return to these ramifications in Sec.~\ref{par:infinitecouplingsmallflavours}.\\

In this section we have proved that the chiral ring relations are all implied by
\begin{equation}
    I\otimes \tilde{I}^{\rm c}=S^{k+1}\ee^{M/S}
\end{equation}
whether $S$ is equal to zero or not. Thus, while the independent degrees of freedom in the IR theory are the mesons $M_{ij}$ and the gaugino bilinear $S$, in the UV the low energy theory on a generic point of the Higgs branch only depends on the instantons.
Furthermore, we have shown how the relative “alignment" of the instantons as in Eq.~\ref{eq:IIt-pluses}, indexed by the type $t$ of the bispinor, labels different orbits of $SO(2N_f)$. We will return to this derivation in Sec.~\ref{sec:stratification}, where we will recover the infinite coupling Hasse diagram for the theories at hand.

\section{Integrability}
\label{sec:integrability}
The vacuum structure of 5d Higgs branches has been studied in the paradigm of \textit{magnetic quivers} \cite{cabrera2019tropical}. Their integrable structure, however, has scarcely been explored so far, and might lead to further physical information on the BPS sector. Our aim here is to introduce the study of integrable structures on the 5d Higgs branch.

The aim of this section is to characterize the symplectic holomorphic two-form from integrability arguments. We will consider some prerequisites of describing the Higgs branch as an algebraically complete integrable system, followed by the application of the Liouville--Arnold theorem and its Darboux corollary, which will allow us to write $\omega$ locally in terms of action-angle variables.

\subsection{Poisson brackets}

\paragraph{Ansatz for the bracket.} 

Recall that, given two functions $f$ and $g$ on the Higgs branch, we defined their Poisson brackets as in Eq.~\ref{eq:generalpoisson}: 
\begin{equation}
    \{f,g\}=(\omega^{-1})^{ij}\partial_i f\partial_j g \,.
    \label{eq:poisson}
\end{equation}
As detailed in \cite{gledhill2023poisson}, the Poisson brackets of chiral ring generators can be bootstrapped without the exact knowledge of $\omega$, allowing to find the right-hand side of Eq.~\ref{eq:poisson} up to numerical coefficients. The bracket of two chiral ring elements of R-charge, respectively, $R_f$ and $R_g$ has to have R-charge:
\begin{equation}\label{eq:Rcharge}
    R_{\{f,g\}}=R_f+R_g-2 \ .
\end{equation}
Furthermore, Poisson brackets must be covariant with respect to the global symmetries under which $f$ and $g$ are charged. The bracket of moment maps of a symmetry group $G$ realize the corresponding Lie algebra. We see from Eq.~\ref{eq:Rcharge} that they have R-charge two. Mesons are the moment maps of the $SO(2N_f)$ symmetry, and thus satisfy:
\begin{subequations}\label{eq:allpoisson}
\begin{equation}
    \{ M_{ij},M_{kl}\}= -4 \delta_{k[i} M_{j]l}+4 \delta_{l[i} M_{j]k} = f_{[ij][kl]}^{[mn]}M_{mn} \,,
    \label{eq:poissonM}
\end{equation}
with $f$ the structure constants of $SO(2N_f)$. Similarly, the Poisson bracket of the moment map of the topological $U(1)$ symmetry, namely the gaugino bilinear $S$, reads:
\begin{equation}
\begin{split}
    \{S,S\}=0 \,,\\
    \{S,M_{ij}\}=0\,.
\end{split}  
\end{equation}
The bracket of a moment map on other operators is similarly determined by the requirement that it should act in the appropriate $G$ representation. Hence:
\begin{equation}
\begin{split}
     \{M_{ij},I \}  & = -\frac14 \gamma_{ij} I \,, \\
     \{S,I\}  & = I\,.
\end{split}
\end{equation}
and similarly for $\tilde{I}$. In this case, the right-hand side is determined to be linear in $I$ also because there are no other operators with $U(1)$ charge $\pm 1$. 

Furthermore, the Poisson brackets of the instantons with themselves have to vanish:
\begin{equation}
    \begin{array}{ll}
    \{I,I\} & =0\,, \\
    \{\tilde{I},\tilde{I}\} & =0 \,.\\
\end{array} \label{eq:poissoncommutation}
\end{equation}
Indeed, the only charge $\pm 2$ element that could fit the right-hand side of Eq.~\ref{eq:poissoncommutation} is in the second symmetric product of the instantons, whose R-charge does not satisfy Eq.~\ref{eq:Rcharge}. The Poisson bracket of the instanton with the anti-instanton is not fully determined by symmetry principles, but  we can write an Ansatz for it compatible with Eq.~\ref{eq:Rcharge}:
\begin{equation}\label{eq:P-II}
	\{I,\tilde I^{\rm c}\} = S^k 1 + S^{k-1} M + S^{k-2} M \wedge M +\ldots + \wedge^{k} M \,\
\end{equation}
\end{subequations}
up to numerical coefficients.
As usual, on the right-hand side we are identifying forms with the corresponding bispinors under the Clifford map Eq.~\ref{eq:clifford}. Eq.~\ref{eq:P-II} is reminiscent of $\mathcal{W}$ algebras, given the non-linear structure of its right-hand side.\\

We can check the Ansatz of Eq.~\ref{eq:P-II} in the case of $k=1$. For $N_f\leq7$, the global symmetry enhances from $SO(2N_f)\times U(1)$ to $E_{N_f+1}$: all the chiral ring generators $M,S,I$ and $\tilde{I}$ have R-charge 2 and conspire to form the adjoint representation of $E_{N_f+1}$. We can branch its adjoint representation in terms of the symmetry at finite coupling $SO(2N_f)\times U(1)$ and recover Eq.~\ref{eq:P-II}.

\paragraph{Compatibility with relations.}

In general, for a bracket to restrict to a submanifold defined by constraints, $\{c_I=0\}$, one needs to check that it is compatible with them. Namely, the condition $\{c_I,f\}=0$ has to be imposed, so that the outcome of the bracket depends only on the value of the functions on $\{c_I=0\}$. Notice that this also implies that Hamiltonian vector fields are tangent to $\{c_I=0\}$, even if generated by Hamiltonians that are functions of the full space: indeed $\delta c_I = \{H,c_I\}=0$.

In our case, we need to check the compatibility of Eq.~\ref{eq:allpoisson} with relations Eq.~\ref{eq:rel}. As we discussed, Eq.~\ref{eq:II} (in particular, for $q=0,2$) generically implies all the others. The most challenging computation is to show 
\begin{equation}
    \{\tilde I\cdot \gamma_{ij}I, \tilde I\cdot \gamma_{lm}I\}=
    S^{2k}\{M_{ij},M_{lm}\}\,. 
\end{equation}
We now sketch an argument to this effect. Using Eq.~\ref{eq:P-II} on left-hand side produces the bilinear $B_{ij,lm}\equiv\tilde I \cdot \gamma_{ij}(S^k + \ldots + \wedge^k M) \gamma_{lm} I - (i\leftrightarrow l, j\leftrightarrow m)$. This can be rewritten using the identity
\begin{equation}
    \stackrel\to {\gamma_{ij}} \stackrel\leftarrow {\gamma^{lm}} 
    = 4\dd x_{[i} \wedge \iota_{j]} (\dd x^l \wedge \dd x^m - \iota^l \iota^m)-4 (\dd x_i \wedge \dd x_j - \iota_i \iota_j)\dd x^{[l} \wedge \iota^{m]}
\end{equation}
which follows from Eq.~\ref{eq:gG}. This operator maps a $2q$-form to $2(q+1)$- and $2(q-1)$-forms. The bilinear $B_{ij,lm}$ will now have $q\pm 1$ copies of $M$ after using $\tilde I\cdot \gamma_{2q} I \propto S^{k+1-q}\wedge^q M$ from Eq.~\ref{eq:II}, and $q$ powers of the coefficients $M_{ij}$ in $M= \frac12 M_{ij} \dd x^i \wedge \dd x^j$, for a total of $2q\pm 1$ powers of $M_{ij}$. After using $M^2 \propto S^2 1$ from Eq.~\ref{eq:M2}, this will reduce to a single power of $M$; the only structure allowed by antisymmetry in $ij$ and $lm$ is then $M_{[i}{}^{[l}\delta_{j]}^{m]}$. This is indeed of the form expected in Eq.~\ref{eq:poissonM}.\footnote{Obtaining the overall coefficient with this argument is challenging for general $N_f$ and $k$; it would be more practical to compute directly the outcome on two particular spinors in the oscillator basis Eq.~\ref{eq:basiz}. In any case, the outcome is not sufficient to fix the coefficients in Eq.~\ref{eq:P-II}, so we have refrained from carrying this out.}

\subsection{Liouville--Arnold theorem}

In this subsection we show that our Higgs branches satisfy all requirements of \emph{algebraically completely integrable systems} (defined in \cite{donagi1995spectral}), which are in turn guaranteed by the Liouville--Arnold theorem. \\
Essentially, integrability means that there are as many functions that Poisson-commute with themselves as half the dimension of the space. Indeed on our Higgs branch the $I$ (or the $\tilde I$) commute, as per Eq.~\ref{eq:poissoncommutation}. 
Moreover, the number of independent $I$ is indeed half the dimension. As mentioned before, the last relation in Tab.~\ref{tab:infinitecouplingconstraintsGENERAL} implies that the bare instantons operators are pure spinors. In particular, the moduli space of one pure spinor in $2N_f$ Euclidean dimensions corresponds to Eq.~\ref{eq:coset}. This space, which is a Gorenstein singularity and has been amply studied \cite{cederwall2012geometry, nekrasov2005lectures}, has half the dimension of the Higgs branch of $Sp(k)$ with $N_f \leq 2k+3$ at strong coupling \cite{bourget2020magnetic}. Indeed, in the range considered, the complex dimension of the Higgs branch is $N_f(N_f-1)+2$. It follows that the space of one pure spinor is a \emph{Lagrangian submanifold} of the Higgs branch: a submanifold of half the dimension, such that the restriction of the Poisson structure to it vanishes. This condition is furthermore guaranteed by the following theorem \cite{arnol2013mathematical, donagi1995spectral}.

\begin{theorem}[Liouville--Arnold]
   Let $M$ be an $m$-dimensional Poisson manifold with Poisson structure $\psi$ 
of constant rank $2g$. Suppose that 
\[
H : M \to B
\]
is a proper submersive Hamiltonian map of maximal rank, i.e.,
\[
\dim B = m - g.
\]
Then:

\begin{enumerate}
    \item The null foliation of $M$ is induced locally by a foliation of $B$ 
    (globally if $H$ has connected fibers).
    
    \item The connected components of the fibers of $H$ are Lagrangian compact 
    tori with a natural affine structure.
    
    \item The Hamiltonian vector fields of the pullback of functions on $B$ 
    by $H$ are tangent to the level tori and are linear in their coordinates.
\end{enumerate}
\end{theorem}

Before applying the theorem to our case, let us point out aspects of the terminology. A proper map is such that the preimage of every compact set is itself compact. It is submersive if differential is surjective at every point. Furthermore, $H$ is Hamiltonian if $\{H^*f,H^g\}=0$ for any $f$, $g$. Moreover, the kernel:
\[
\text{ker}(\omega|_N)=\{v \in T_PN\,|\, \omega(v,v')=0 \,\text{ for all }v'\in T_pN\}
\]
is called the null foliation of $M$. It represents the locus where $\omega$ degenerates. In the context of the Higgs branches we are considering, we need to consider the holomorphic equivalent \cite{donagi1995spectral} of the Liouville--Arnold theorem, where $\omega$ is now the holomorphic symplectic 2-form; furthermore, we need to consider as $M$ the largest leaf of the Higgs branch, which is by definition a smooth space with a Poisson structure of constant rank and has the same dimension as the Higgs branch.
In the holomorphic version, the Lagrangian tori become abelian varieties, which are complex tori, namely quotients of complex flat space by a lattice, that are projective, i.e.~admit an embedding in complex projective space. 
\\

 A natural candidate for the submanifold $B$ is the cone $Q$ of a single pure spinor, which is half dimensional. The submersive maps that parameterize it correspond to the subset of the components of the pure spinor obtained by branching under the $U(N_f)$ subgroup:
 \begin{equation}\label{eq:branching}
     I_{\alpha}\overset{U(N_f)}{\longrightarrow}(\underset{I_i}{\underbrace{\lambda, \lambda_{[ab]}}},\dots ) \qquad \text{with}\quad a,b=1,\dots N_f \ ,
 \end{equation}
 In particular, in compliance with half dimensionality, there are $N_f(N_f-1)/2+1$ complex maps. These are of maximal rank as they are independent; furthermore, since they Poisson-commute with themselves, they are Hamiltonian maps.\\
 Out of these, only $N_f(N_f-1)/2$ of them are proper, namely $\lambda_{[ab]}$ in Eq. ~\ref{eq:branching}, since their image is the coset $SO(2N_f)/U(N_f)$. One map is not proper, $\lambda$, since it runs along  the conical direction. This implies \cite{donagi1995spectral, donagi1996supersymmetric} that the fibers are \textit{generalised tori}. Concretely, the fibers are isomorphic to a $N_f(N_f-1)/2$ dimensional complex torus times a complex line:  $T^{N_f(N_f-1)/2}\times \mathbb{C} $. \\

 As a consequence of the theorem by Liouville--Arnold, we can claim that the Higgs branch, on its largest leaf, is an \textit{algebraically complete integrable Hamiltonian system}. Its mathematical definition can be given in algebraic-geometric language \cite{donagi1995spectral, nekrasov2009quantization}:
\begin{definition}[Algebraically Completely Integrable Hamiltonian System]
An algebraically completely integrable Hamiltonian system consists of a 
proper flat morphism
\[
H : M \to B,
\]
where $(M,\psi)$ is a smooth Poisson variety and $B$ is a smooth variety, 
such that, over the complement $B \setminus \Delta$ of some proper closed 
subvariety $\Delta \subset B$, the map $H$ is a Lagrangian fibration whose 
fibers are isomorphic to abelian varieties.
\end{definition}

The \emph{variety} is the locus cut out by the chiral ring equations.  
In particular, the morphism $H$ is a map compatible with the Poisson structure on M and its flatness ensures continuity of the fibers over $\Delta$.

\subsection{Action-angle variables}\label{sec:actionangle}
As a corollary to the Liouville--Arnold theorem, due to Darboux, we can find local coordinates on the manifold $B$, called action coordinates $I_i$, and angular coordinates on the torus, named angle coordinates $\phi_i$. Furthermore, the holomorphic symplectic structure can be written locally as:
\begin{equation}
    \omega=\sum_{i=1}^{N_f(N_f-1)/2+1} \text{d}I_i \wedge \text{d}\phi_i 
\end{equation}
in terms of the $I_i$ defined in Eq.~\ref{eq:branching}.
Since the Higgs branch has scaling symmetry, it cannot have global one-cycles, hence the toric fibration degenerates at least at the origin of the space. \\

 The question may arise whether we should consider the instantons or the anti-instantons as action coordinates, as they both are pure spinors and satisfy all criteria. We can equally as well write $\omega$ locally in terms of the anti-instanton:
\begin{equation}
    \omega= \sum_{i=1}^{N_f(N_f-1)/2+1} \text{d}\tilde{I}_i \wedge \text{d}\tilde{\phi}_i \,.
\end{equation}
The two expressions are only valid in a given coordinate patch. Already on the space of one pure spinor, there does not exist a globally defined set of coordinates, as discussed in \cite{nekrasov2005lectures}, but at least $2^{N_f-1}$ patches. On the Higgs branch, furthermore, the situation is worsened by the would-be union of two cones structure, discussed in section \ref{sec:chiralring}.
We can define a projector on each cone: in terms of the bispinor $\Phi$ in Eq.~\ref{eq:Phi}, for $S\neq 0$ we can write 
\begin{equation}
I =S^{-k}\Phi I \,.
\end{equation}
As described in the previous section, the bispinor $\Phi$ is a pure form and has an annihilator of at most $2N_f$ dimensions, or lower in case of alignment of pure spinors. Furthermore, if the spinors align along some directions, or equivalently if their respective annihilators overlap, the bispinor $\Phi$ takes values on different orbits under $SO(2N_f)$.\\

In this section, we have described the 5d Higgs branch as the phase space of an integrable system, or equivalently as the local fibration of tori on the union of two cones. In particular, we claim that the degeneration of the fibration is determined by the relative alignment of the pure spinors, namely by the varying annihilator of the projector $\Phi$ in Eq.~\ref{eq:Phi}. As a consequence, the symplectic leaves are determined by the bispinor orbits. We summarize the correspondence in Table \ref{tab:integrable}.
\begin{table}[ht]
    \centering
    \begin{tabular}{|ccc|} \hline
       Gauge theory picture  &  &Integrable picture \\ \hline \hline
       Higgs branch variety   & $\longleftrightarrow$ &  Phase space \\
       \makecell{vevs of instanton\\ operators}   & $\longleftrightarrow$ & Action variables \\
       Symplectic leaves  & $\longleftrightarrow$ & Orbits of projector \\ \hline
    \end{tabular}
    \label{tab:integrable}
    \caption{Claim of integrable interpretation of the Higgs branch at study.}
\end{table}

This point of view is put to test in the next section, where we recover the stratification of the Higgs branches in question, both at finite and infinite coupling.

\section{Stratification and bispinor orbits}
\label{sec:stratification}

The claim at the end of the last section was that the stratification of the Higgs branch, at weak or strong coupling, is determined by the orbits of the bispinor.
We will now ground this idea by explicitly building the stratification out of the bispinor orbits; its diagrammatic representation is given by the Hasse diagram, whose physical meaning we summarise in the following paragraph. We thus proceed to recover the Hasse diagram for $k=1$ and then generalise to arbitrary $k$ and $N_f$, in the regime we are considering of $N_f\leq 2k+3$.

\paragraph{Hasse diagram basics.}
\label{sub:hasse}

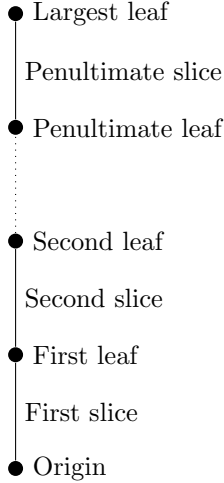
\begin{figure}[ht]
\centering

\begin{minipage}{0.45\textwidth}
    \centering
  \begin{tikzpicture}[
    node distance=1.3cm,
    dot/.style={circle, fill, inner sep=2pt},
    every edge quotes/.style={auto, font=\small}
]

\node[dot, label=right:{Origin}] (L1) {};
\node[dot, above=of L1, label=right:{First leaf}] (L2) {};
\node[dot, above=of L2, label=right:{Second leaf}] (L3) {};
\node[dot, above=of L3, label=right:{Penultimate leaf}] (L5) {};
\node[dot,above=of L5, label=right:{Largest leaf}] (L4) {};

\draw (L1) -- node[right] {First slice}  (L2);
\draw (L2) -- node[right] {Second slice}  (L3);
\draw[dotted] (L3) -- node[right] {} (L5);
\draw (L5) -- node[right] {Penultimate slice}  (L4);
\end{tikzpicture}
\end{minipage}
\caption{General form of a linear Hasse diagram.}
\label{fig:generalHasse}
\end{figure}

For our purposes, the Hasse diagram will be a representation of the stratification of the Higgs branch in symplectic leaves (dots), connected by slices (edges), as in Fig.~\ref{fig:generalHasse}.

As we mentioned earlier, a Poisson manifold can be foliated in symplectic leaves according to the rank of the Poisson tensor, namely the inverse of the holomorphic symplectic two-form. Each leaf $l$ is a symplectic manifold, with dimension equal to the rank of the Poisson tensor $P$ at every point on $l$. In other words, at every point the image of $P$ is the tangent space to $l$. An equivalent way to phrase this is that each leaf is closed under Hamiltonian deformations $\delta x = \{H,x\}$, which indeed can be written as $\delta x^i= P^{ij}\partial_j H$.  This point of view will be useful to us later.\\
The largest leaf of the Higgs branch is an open set thereof and has its same dimensions, while the lower leaves 
have lower dimensions and lower rank of $\omega$, and are obtained by tuning some of the moduli to zero. The origin corresponds to the point where all moduli are set to zero, namely the origin of the whole moduli space of vacua. \\

The slices are orthogonal directions to the leaves. To understand this in a simple, non holomorphic case, let us consider the well known Mexican hat potential: in that case the moduli space of vacua would be the circle of degenerate vacua, while the slice would be the transverse, radial direction across the slopes of the potential. More generally, the slice between two leaves can be seen as the space of moduli we have to set to zero in order to restrict the larger leaf to the smaller one.\\
In \cite{bourget2020higgs}, all details are provided to compute the slices and leaves of Hasse diagrams of a wide class of theories. Furthermore, much progress has been achieved in recent years in characterizing some slices as Coulomb branches of some 3d $\mathcal{N}=4$ gauge theories, or equivalently \cite{cremonesi2014coulomb} as moduli space of one $G$-instanton on $\mathbb{C}^2$, namely as \textit{minimal nilpotent orbits}.\\

Any sub-diagram of the Hasse diagram is a Hasse diagram itself and may identify the Higgs branch of a potentially different 5d theory, as we are going to point out when convenient. Furthermore, leaves may correspond to the opening of a mixed Higgs branch -- Coulomb branch direction. This is clear to see in Hasse diagrams of Lagrangian theories: while the origin corresponds to unbroken gauge group, higher leaves correspond to partial or complete breaking. The largest leaf corresponds (if enough matter is present) to complete breaking of the gauge group and is populated by the residual interacting massless flavours. The penultimate leaf corresponds to partial symmetry restoration, hence a Coulomb branch direction opens up, corresponding to a massless vector multiplet. Similarly for lower leaves, until we reach the origin of the moduli space from which the full $k$ dimension Coulomb branch departs, where $k$ is the rank of the gauge group.\\

In strongly coupled theories the notion of gauge symmetry is ill-defined, nonetheless we can still argue that at least part of the symplectic leaf corresponds to mixed branches. In particular, at strong coupling, the Coulomb branch of the theories we are considering is still $k$-dimensional, as no known effects produce an increase in dimensions, unlike the appearance of massless instantons on Higgs branches.

\subsection{\texorpdfstring{Hasse diagram for $Sp(1)$}{Hasse diagram for Sp(1)}}\label{sec:hassediagramSp(1)}
In this section we will consider the simplest example: $Sp(1)$ with $N_f \leq 7$ flavours, at weak and strong coupling. This will be useful in identifying slices of more general Higgs branches; namely every subdiagram of a Hasse diagram can be identified with the Hasse diagram of a different theory's Higgs branch. In the present case, we will identify slices between adjacent leaves of Hasse diagrams of $Sp(k)$ theories as Higgs branches of $Sp(1)$, at either weak or strong coupling.\\
 In both limits, the Higgs branch \cite{cremonesi2017instanton} is a minimal nilpotent orbit, the difference being between $d_{N_f}$ orbits at finite coupling and $E_{N_f+1}$ orbits at infinite, following the Seiberg symmetry enhancement.
\begin{figure}[ht]
\centering

\begin{minipage}{0.45\textwidth}
    \centering
    \textbf{Weak coupling}
    
    
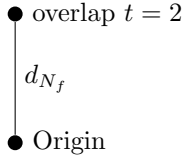
\vspace{0.5cm}
    
      \begin{tikzpicture}[
    node distance=1.5cm,
    dot/.style={circle, fill, inner sep=2pt},
    every edge quotes/.style={auto, font=\small}
]

\node[dot, label=right:{Origin}] (L1) {};
\node[dot, above=of L1, label=right:{overlap $t=2$}] (L2) {};

\draw (L1) -- node[right] {$d_{N_f}$}  (L2);
\end{tikzpicture}
\end{minipage}
\hfill
\begin{minipage}{0.45\textwidth}
    \centering
    \textbf{Strong coupling}
    
    
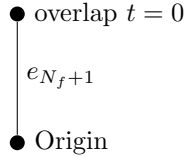
\vspace{0.5cm}
    
     \begin{tikzpicture}[
    node distance=1.5cm,
    dot/.style={circle, fill, inner sep=2pt},
    every edge quotes/.style={auto, font=\small}
]

\node[dot, label=right:{Origin}] (L1) {}; 
\node[dot, above=of L1, label=right:{overlap $t=0$}] (L2) {};

\draw (L1) -- node[right] {$e_{N_f+1}$}  (L2);
\end{tikzpicture}
\end{minipage}

\end{figure}

Being a minimal nilpotent orbit, the Hasse diagram contains only one slice. The Higgs branch at infinite coupling is determined by the chiral ring first examined in \cite{cremonesi2017instanton} and extended at higher rank in \cite{hanany2025chiral}. As we argued in Sec.~\ref{sec:chiralring}, the generic case is 
$I\otimes\tilde{I}=S^2 \exp(M/S)$,  a pure form of type $t=0$, with two disjoint annihilators: ${\rm Ann}(I)\cap {\rm Ann}(\tilde I)=0$. However, in the generic case, the two pure spinors have 
\begin{equation}\label{eq:purespinorfock}
\tilde{I} =|+^{N_f}\rangle \,,\qquad
    I =|-^{N_f}\rangle + \ldots \,,
\end{equation}
where $\ldots$ denotes spinors with some plus signs.
Using the results of \cite{cremonesi2017instanton}, we can write
\[
\boxed{
e_{N_f+1} \quad \supset \quad \begin{array}{l}
     \text{Moduli space of two disjoint isotropic $N_f$-dim subspaces}  \\
     \text{ in a $2N_f$-dim euclidean space.}
\end{array}}
\]
with the right-hand side being dense.

At weak coupling, we know the instantons to be massive and thus decoupled from the chiral ring. In a sense, however, we can still use them to characterize the stratification: we can consider the instantons that would be associated by the mesons by Eq.~\ref{eq:MI}, were it not for the other relations. For example, for $M = \dd z^1 \wedge \dd z^2$, in the orbit $[2^2,1^{N_f-4}]$, the generic spinor pair would read
\begin{equation}\label{eq:purespinorfock2}
\tilde{I} =|+^{N_f}\rangle \,,\qquad
    I =|+^2-^{N_f}\rangle + \ldots \,,
\end{equation}
with the dots again denoting a lower number of minus signs. A rank-two meson matrix indeed parameterises the leaf of $Sp(1)$ at weak coupling. \\
We can identify the minimal nilpotent orbit of $d_{N_f}$ as the moduli space of a 2 dimensional subspace in a $2N_f$ dimensional isotropic euclidean space:
\[
\boxed{
d_{N_f} \quad = \quad \begin{array}{l}
     \text{Moduli space of a 2-dim subspace}  \\
     \text{in a $2N_f$-dim isotropic euclidean space.}
\end{array}}
\]
The projectivized part of this moduli space is already known in the literature \cite{collingwood1993nilpotent} and goes under the name of orthogonal Grassmannian $\text{OGr}(2,2N_f)$, which has $\text{dim}_{\mathbb{C}}=4N_f-7$. The Orthogonal Grassmanian can be written as the coset:
\begin{equation}
   \text{OGr}(2,2N_f)= \dfrac{SO(2N_f)}{SO(2N_f-4)\times U(2)} \ ,
\end{equation}
upon which $d_{N_f}$ is the complex cone.

Before we move on to more complicated examples, let us mention the degenerate case of $e_{N_f+1}$ with $N_f=0$ and $N_f=1$. In particular, $e_1$ corresponds to a Klein singularity $A_1$, i.e.~$\mathbb{C}^2/\mathbb{Z}_2$. The minimal nilpotent orbit $e_2$ also corresponds to the same Klein singularity, except for a discrete part of the variety. Klein singularities are the degenerate case of exceptional nilpotent orbits in the sense that, according to the definition above, they correspond to the moduli space of two disjoint zero (respectively, one) dimensional subspaces in a zero (or two) dimensional euclidean space. We thus expect them to arise when all $N_f$ (respectively $N_f-1$) directions of the annihilator overlap in Eq.~\ref{eq:purespinorfock}.

\subsection{General Hasse diagram}

We will now reproduce the structure of the Hasse diagram for $N_f\leq 2k+3$, excluding the cases $N_f\in \{2k+4,2k+5\}$. 

\begin{figure}[ht]
\centering
\begin{minipage}{0.45\textwidth}
\centering
\begin{tikzpicture}[
    node distance=2cm,
    dot/.style={circle, fill, inner sep=2pt},
    every edge quotes/.style={auto, font=\small}
]
\node[dot, label=right:{$M\in[1^{2N_f}]$}] (L1) {};
\node[dot, above=of L1, label=right:{$M\in[2^2,1^{2(N_f-2)}]$}] (L2) {};
\node[dot, above= of L2, label=right:{$M\in[2^{2(k-1)},1^{2(N_f-2k+2)}]$}] (L3) {};
\node[dot, above= of L3, label=right:{$M\in[2^{2k},1^{2(N_f-2k)}]$}] (L4) {};

\draw (L1) -- node[right] {$d_{N_f}$} (L2);
\draw[dotted] (L3) -- (L2);
\draw (L4) -- node[right] {$d_{N_f-2k+2}$} (L3);
\end{tikzpicture}
\end{minipage}
\caption{Weak-coupling Hasse diagram for $2k+1\leq N_f\leq 2k+3$.}
\label{fig:hasse-M2}
\end{figure}
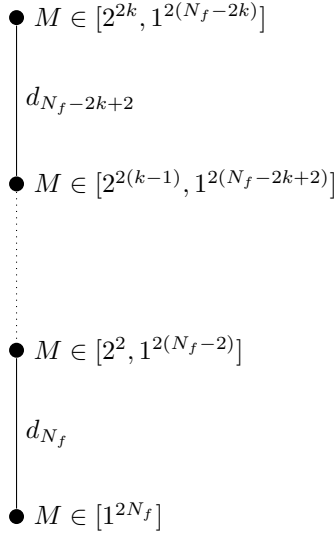

\paragraph{Weak coupling case.} We begin with the weak-coupling case, whose relations were given in Eq.~\ref{eq:w-rel}. The only non-zero Poisson bracket is Eq.~\ref{eq:poissonM}, of $M$ with itself. Recall from Sec.~\ref{sub:hasse} that a symplectic leaf is closed under Hamiltonian deformations; the only non-trivial choice is 
\begin{equation}\label{eq:HM}
    H=a^{ij} M_{ij}   \,,
\end{equation}
which gives $\delta M = \{H,M\} \propto [a,M]$, the matrix commutator. So at every point $M$ the tangent space to the leaf is given by the orbit of $M$ under the adjoint action of $\mathfrak{so}(2N)$. Since $M$ is nilpotent, we are left with the Hasse diagram of $\mathfrak{so}(2N)$ nilpotent orbits. The leaves $l^{\rm w}_i$ can be labeled by $i=\mathrm{rank}M/2$, and the corresponding orbit is $[2^{2i},1^{2(N_f-2i)}]$ as we saw in Eq.~\ref{eq:M-orbits}; the largest rank is ${\rm min}(2k,N_f)$. The resulting Hasse diagrams are shown in Fig.~\ref{fig:hasse-M}, \ref{fig:hasse-M2}.

\begin{figure}[ht]
\centering
\begin{minipage}{0.45\textwidth}
\centering
\begin{tikzpicture}[
    node distance=2cm,
    dot/.style={circle, fill, inner sep=2pt},
    every edge quotes/.style={auto, font=\small}
]

\node[dot, label=right:{$M\in[1^{2N_f}]$}] (L1) {};
\node[dot, above=of L1, label=right:{$M\in[2^2,1^{2(N_f-2)}]$}] (L2) {};
\node[dot, above= of L2, label=right:{$M\in[2^{N_f-2},1^4]$}] (L3) {};
\node[dot, above= of L3, label=right:{$M\in[2^{N_f}]$}] (L4) {};

\draw (L1) -- node[right] {$d_{N_f}$} (L2);
\draw[dotted] (L3) -- (L2);
\draw (L4) -- node[right] {$d_2$} (L3);
\end{tikzpicture}
\end{minipage}
\hfill
\begin{minipage}{0.45\textwidth}
\centering
\begin{tikzpicture}[
    node distance=2cm,
    dot/.style={circle, fill, inner sep=2pt},
    every edge quotes/.style={auto, font=\small}
]

\node[dot, label=right:{$M\in[1^{2N_f}]$}] (L1) {};
\node[dot, above=of L1, label=right:{$M\in[2^2,1^{2(N_f-2)}]$}] (L2) {};
\node[dot, above= of L2, label=right:{$M\in[2^{N_f-3},1^6]$}] (L3) {};
\node[dot, above= of L3, label=right:{$M\in[2^{N_f-1},1^2]$}] (L4) {};

\draw (L1) -- node[right] {$d_{N_f}$} (L2);
\draw[dotted] (L3) -- (L2);
\draw (L4) -- node[right] {$d_3$} (L3);
\end{tikzpicture}
\end{minipage}
\caption{Weak-coupling Hasse diagram for $N_f\leq 2k$ and $N_f$ even (left) or odd (right), respectively. The order of the slice changes by two units as we go up the diagram: from $d_{N_f}$ to $d_{N_f-2}$ and so forth. Furthermore, $d_2$ can be considered as the product $a_1\times a_1$, hence showing that the Higgs branch is the union of two cones.}
\label{fig:hasse-M}
\end{figure}
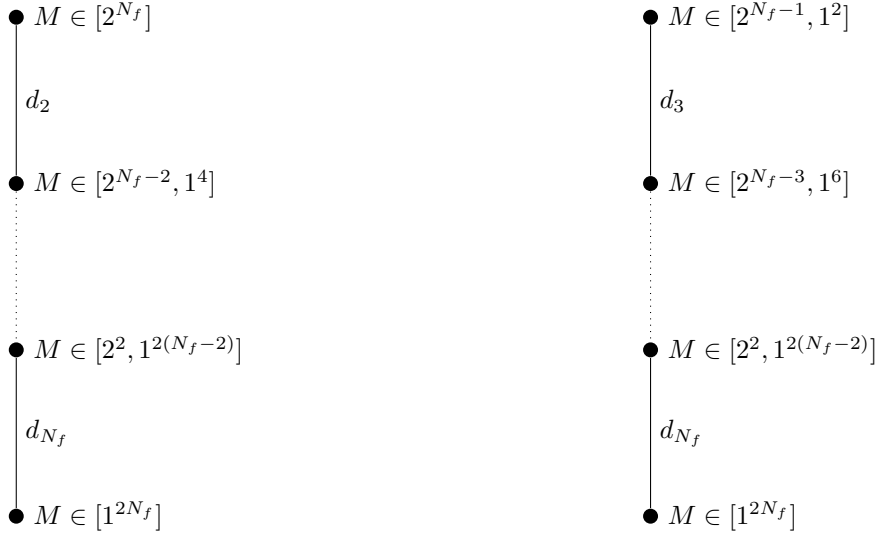

\medskip

 The infinite coupling case is much more involved. We can still use Eq.~\ref{eq:HM}; it now also acts on the instantons, so every leaf is a union of orbits for the action of $\mathfrak{so}(2N)$ on the full set of data $(M,S,I,\tilde I)$. But we can now also use Hamiltonians that are functions of instantons, such as
 \begin{equation}\label{eq:HI}
     H= \eta \cdot I + \tilde \eta \cdot \tilde I\,.
 \end{equation}
As we shall see, this sometimes means that a leaf is made of many $\mathfrak{so}(2N)$ orbits.  We divide the discussion into two cases.

\paragraph{Infinite coupling: $N_f\leq 2k-1$.}\label{par:infinitecouplingsmallflavours}

In this case, all the symplectic leaves $l^{\rm w}_i$ that we saw for finite coupling become symplectic leaves at infinite coupling, if we also set $S=I=\tilde I=0$. 

 First, the locus $\{S=I=\tilde I=0\}$ is trivially invariant under Eq.~\ref{eq:HM}: $\delta I = \{H,I\}= a^{ij}\gamma_{ij}I=0$, and similarly $\delta S=\delta \tilde I=0$. Moreover, $\{S=I=\tilde I=0\}$ is also closed under Hamiltonian flows Eq.~\ref{eq:HI}. Indeed, $\wedge^k M=0$, so the right-hand side of Eq.~\ref{eq:P-II} vanishes: $\{I,\tilde I\}=0$. It then follows $\delta I = \{ H, I \}=0$, and $\delta S$, $\delta \tilde I$ vanish as well. All in all we obtain infinite-coupling leaves
\begin{equation}
    l^\infty_i = l^{\rm w}_i \cap \{S=I=\tilde I=0\}\qquad i=1,\ldots,\lfloor N_f/2 \rfloor\,.
\end{equation}

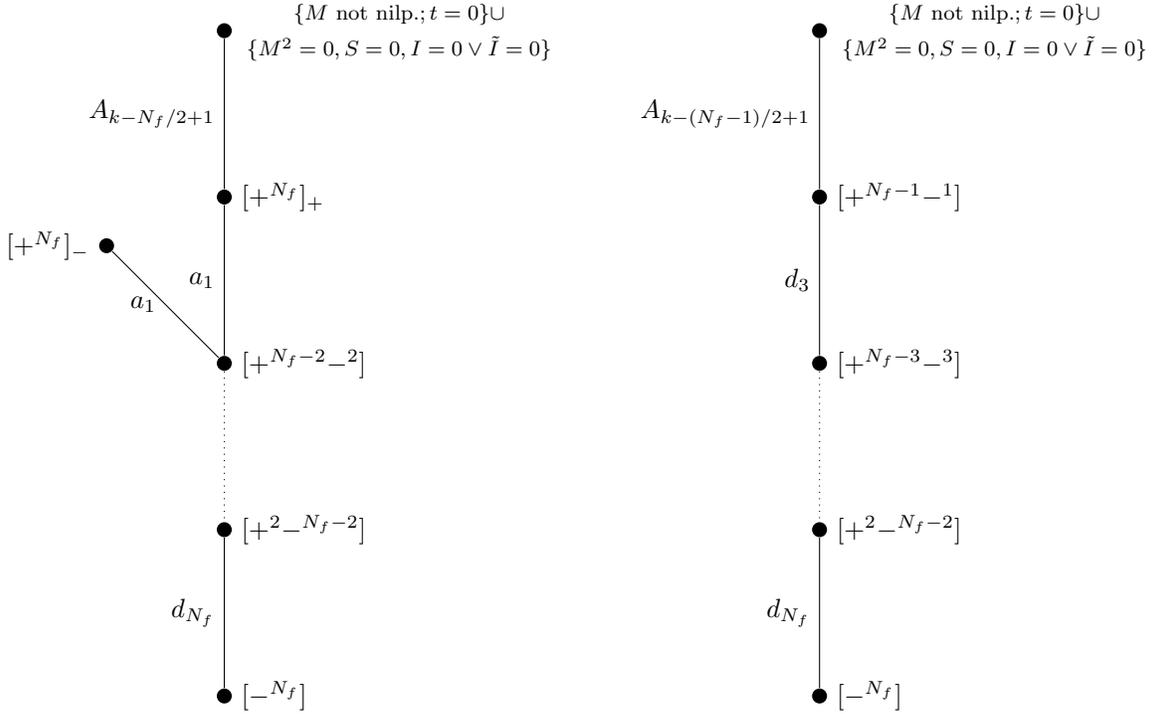
\begin{figure}[ht] 
\centering
\begin{minipage}{0.45\textwidth}
\centering
\begin{tikzpicture}[
    node distance=2cm,
    dot/.style={circle, fill, inner sep=2pt},
    every edge quotes/.style={auto, font=\small}
]

\node[dot, label=right:{$[-^{N_f}]$}] (L1) {};
\node[dot, above=of L1, label=right:{$[+^2-^{N_f-2}]$}] (L2) {};
\node[dot, above= of L2, label=right:{ $[+^{N_f-2}-^{2}]$}] (L3) {};
\node[dot, above= of L3, label=right:{$[+^{N_f}]_+$}] (L4) {};
\node[dot, above= of L4, label=right:{\footnotesize $\begin{array}{c}\{M\text{ not nilp.}; t=0\}\cup \\
\{M^2=0, S=0, I=0 \lor \tilde I = 0 \}\end{array}$}] (L5) {};
\node[dot, above left= of L3, label=left:{$[+^{N_f}]_-$}] (L4bis) {};

\draw (L1) -- node[left] {$d_{N_f}$} (L2);
\draw[dotted] (L3) -- (L2);
\draw (L4) -- node[left] {$a_1$} (L3);
\draw (L4bis) -- node[left] {$a_1$} (L3);
\draw (L5) -- node[left] {$A_{k-N_f/2+1}$} (L4);
\end{tikzpicture}
\end{minipage}
\hfill
\begin{minipage}{0.45\textwidth}
\centering
\begin{tikzpicture}[
    node distance=2cm,
    dot/.style={circle, fill, inner sep=2pt},
    every edge quotes/.style={auto, font=\small}
]
\node[dot, label=right:{$[-^{N_f}]$}] (L1) {};
\node[dot, above=of L1, label=right:{$[+^2-^{N_f-2}]$}] (L2) {};
\node[dot, above= of L2, label=right:{$[+^{N_f-3}-^3]$}] (L3) {};
\node[dot, above= of L3, label=right:{$[+^{N_f-1}-^1]$}] (L4) {};
\node[dot, above= of L4, label=right:{\footnotesize $\begin{array}{c}\{M\text{ not nilp.}; t=0\}\cup \\
\{M^2=0, S=0, I=0 \lor \tilde I = 0 \}\end{array}$}] (L5) {};

\draw (L1) -- node[left] {$d_{N_f}$} (L2);
\draw[dotted] (L3) -- (L2);
\draw (L4) -- node[left] {$d_3$} (L3);
\draw (L5) -- node[left] {$A_{k-(N_f-1)/2+1}$} (L4);
\end{tikzpicture}
\end{minipage}
\caption{Infinite-coupling Hasse diagram for $N_f\leq 2k-1$, for $N_f$ even (left) and odd (right). The choice of starting with $[-^{N_f}]$ or $[+^{N_f}]$ is inconsequential. All leaves, except the top one, have $S=0=I=\tilde{I}$. For $N_f=2k$, the two slices $a_1$ and $A_{k-N_f/2+1}$ degenerate to a single $a_2$, as described in \cite{hanany2025higgs}.}
\label{fig:hasse-Nfsmall}
\end{figure}

We still have to consider the orbits where $I\neq 0$ and/or $I\neq 0$. These form together a single additional leaf. To show this, let us start from $\{I = | +^{N_f} \rangle, \tilde I=M=S=0\}$, and take $H=\eta \cdot I$ (Eq.~\ref{eq:HI} with $\tilde\eta=0$), which generates the flow
\begin{equation}\label{eq:flow-eI}
	\dot I = \partial_\tau I=0  \,,\qquad \dot {\tilde I} = \{\eta \cdot I,\tilde I\}= \sigma(-)^k \wedge^k M  \eta \,,\qquad  \dot M = -\frac12\eta \cdot \gamma_{(2)} I  \,,\qquad \dot S = \eta \cdot  I \,.
\end{equation}
\begin{itemize}
    \item Taking $\eta=|+^2 -^{N_f-2}\rangle$, we see that $\dot{\tilde I}=0$, but $\dot M\propto \langle -^2 +^{N_f-2}| \gamma_{(2)} |+^{N_f}\rangle\sim \dd z^1 \wedge \dd z^2$, of rank $2$. More generally, $\eta=a_1|+^2 -^{N_f-2}\rangle + a_2 |-^2 +^2 +^{N_f-4}\rangle +\ldots$ generates an $M$ of any rank $\le N_f<2k$. No $\tilde I$ is generated along the flow. We reach in this way the leaves $\{S=0,\tilde I=0\}$ in Sec.~\ref{sub:ring}. 
    \item With $\eta=|-^{N_f}\rangle$, we find $\dot S\neq 0$, $\dot M \sim \langle +^{N_f}| \gamma_{(2)} | +^{N_f}\rangle\sim \sum_{a=1}^{N_f} \dd z^a \wedge d \bar z^{\bar a} $. So $M= \tau \sum_a \dd z^a\wedge d \bar z^{\bar a} $. Solving Eq.~\ref{eq:flow-eI} for $\tilde I $  gives $ \sim \tau^{k+1} | -^{N_f} \rangle $.  So the type $t(I,\tilde I)=0$, consistent with $S\neq 0$. We can similarly reach other spinors in this orbit.
\end{itemize}
All in all,
\begin{equation}
    l^\infty_{\lfloor N_f/2 \rfloor+1}= \{S=0,\tilde I=0, M^2=0\} \cup \{S=0, I=0, M^2=0\}\cup \{S\neq 0, I \neq 0, \tilde I \neq 0\}\,.
\end{equation}
In conclusion, in this case we need to add a single leaf to the weak-coupling discussion. We depict the resulting Hasse diagram in Fig.~\ref{fig:hasse-Nfsmall}. We used the notation
\begin{equation}
    [+^{2q} -^{N_f-2q}] \equiv \left\{\begin{array}{c}
         M\in [2^{2q},1^{2(N_f-2q)}]\\
         S=I=\tilde I=0\,.  
    \end{array}\right.
\end{equation}
The idea of the notation is that $M= \dd z^1 \wedge \dd z^2+\ldots + \dd z^{2q-1} \wedge \dd z^{2q}$ (which is in $[2^{2q},1^{2(N_f-2q)}]$) annihilates the spinor $|+^{2q} -^{N_f-2q}\rangle$, as in Eq.~\ref{eq:MI}. The particular case $[2^{2q}]_\pm$ is denoted as $[\pm^{N_f}]$.

The slice from the origin to the first leaf corresponds to the choice of a 2-dim subspace in a $2N_f$-dimensional Euclidean isotropic space, hence the first slice is $d_{N_f}$. After freezing out this 2-dim subspace, we are left with $N_f-2$ directions in the annihilator; choosing 2 of them corresponds to a $d_{N_f-2}$ slice, which brings to a leaf parameterized by the mesons with rank 4. This procedure can be repeated until we reach maximal overlap. The only noticeable difference between even and odd cases is that, for $N_f$ odd, in the penultimate leaf we are left with a single pair of + and $-$ signs in the Fock basis representations of the spinors. This corresponds to a degenerate exceptional orbit, as argued previously in Sec.~\ref{sec:hassediagramSp(1)}, namely a Klein singularity.\\
While we can explain all lower slices, we are currently unable to argue why the top slice should be of type $\mathbb{C}^2/\mathbb{Z}_{h^{\vee}-\lfloor N_f/2\rfloor}$, which is still an open point which will need addressing in future work.\\

For what concerns $N_f$ even, we observe the bifurcation of the Hasse diagram as found by quiver subtraction \cite{bourget2020magnetic}, as anticipated at the end of Sec.~\ref{sub:ring}. Indeed, on the left of Fig.~\ref{fig:hasse-Nfsmall}, the $d_2$ slice corresponds to a product of two $a_1$ slices, leading to orbits of different chirality. As discussed in Sec.~\ref{par:I=0}, one slice leads to the orbit $[2^{N_f}]_{-}$ and the other to $[2^{N_f}]_{+}$. The first orbit implies that $I,\tilde{I}$ vanish on the leaf, while the other orbit allows either $I$ or $\tilde{I}$ to be non-vanishing, hence leading to the observed enhancement.
    
\paragraph{Infinite coupling: $2k \leq N_f \leq 2k+3$.}

In this case, at infinite coupling all of the weak-coupling leaves are unchanged, except the last; we will show that no additional leaf is added.

 Consider first all the weak-coupling leaves with ${\rm rank}M< 2k$, again setting $S=I=\tilde I=0$. As before, we have $\{I,\tilde I\}=0$, and we cannot generate $\delta I$, $\delta \tilde I$, $\delta S\neq 0$, similarly to the previous case:
\begin{equation}
    l^\infty_i = l^{\rm w}_i \cap \{S=I=\tilde I=0\}\qquad i=1,\ldots, k-1 \,.
\end{equation}

 Consider now the weak-coupling leaf $l^{\rm w}_k$, where ${\rm rank} M = 2k$. The locus $l^{\rm w}_k\cap \{S=I=\tilde I=0\}$ is not an infinite-coupling leaf. We will use a first Hamiltonian flow $\dot f \equiv \partial_\tau f = \{H,f\}$ to generate $I$, then a second to also generate $\tilde I$. 

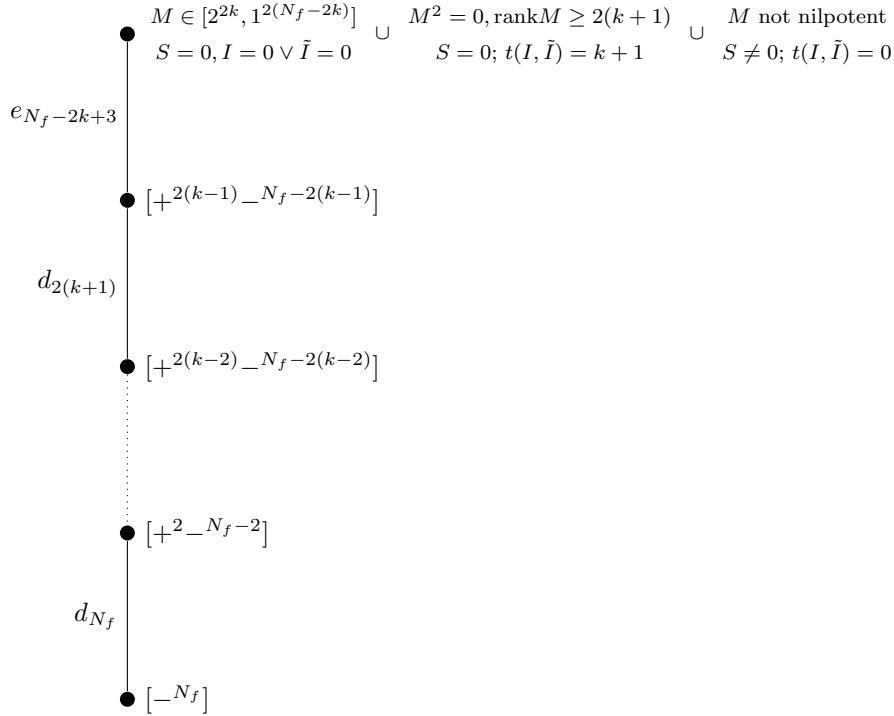
\begin{figure}[htb]
\centering
\begin{tikzpicture}[
    node distance=2cm,
    dot/.style={circle, fill, inner sep=2pt},
    every edge quotes/.style={auto, font=\small}
]

\node[dot, label=right:{$[-^{N_f}]$}] (L1) {};
\node[dot, above=of L1, label=right:{$[+^2-^{N_f-2}]$}] (L2) {};
\node[dot, above= of L2, label=right:{$ [+^{2(k-2)}-^{N_f-2(k-2)}]$}] (L3) {};
\node[dot, above= of L3, label=right:{$ [+^{2(k-1)}-^{N_f-2(k-1)}]$}] (L4) {};
\node[dot, above= of L4, label=right:{
    {\footnotesize$\begin{array}{c}M\in[2^{2k},1^{2(N_f-2k)}]\\
    S=0,I=0 \lor \tilde I=0\end{array}\ \cup \  
    \begin{array}{c}
    M^2=0, {\rm rank}M \geq 2(k+1)\\
    S=0;\, t(I,\tilde I)=k+1
    \end{array}
\ \cup \ 
    \begin{array}{c}M\text{ not nilpotent}\\
    S\neq 0;\, t(I,\tilde I)=0\end{array}$}}] (L5) {};

\draw (L1) -- node[left] {$d_{N_f}$} (L2);
\draw[dotted] (L3) -- (L2);
\draw (L4) -- node[left] {$d_{2(k+1)}$} (L3);
\draw (L5) -- node[left] {$e_{N_f-2k+3}$} (L4);
\end{tikzpicture}
\caption{Infinite-coupling Hasse diagram for $2k\leq N_f\leq 2k+3$. The choice of starting from $[+^{N_f}]$ instead of $[-^{N_f}]$ is inconsequential. All leaves, except the top one, have $S=0=I=\tilde{I}$.}
 \label{fig:hasse-largeNf}
\end{figure}

For the first step, we take $H=\tilde \eta \cdot \tilde I$ (namely, Eq.~\ref{eq:HI} with $\eta=0$). We find
\begin{equation}
	\dot I = \{ \tilde \eta \cdot \tilde I, I \} = - \wedge^k M \tilde\eta \,,\qquad \dot {\tilde I} = 0 \,,\qquad \dot M = -\frac12\tilde \eta \cdot \gamma_{(2)} \tilde I = 0 \,,\qquad \dot S = \tilde \eta \cdot \tilde I = 0 
\end{equation}
Crucially, $\wedge^k M\neq 0$, so $\dot I\neq 0$. Since $M$ is constant along the flow, a solution is simply $I=-\tau \wedge^k M \tilde\eta $. We see that we can generate any $I$ this way. For example, with $M=M_0\equiv \dd z^1 \wedge \dd z^2 + \ldots + \dd z^{2k-1} \wedge \dd z^{2k}$, choosing $\tilde\eta = |-^{2k}+^{N_f-2k}\rangle$ we can reach the canonical choice $I=I_0\equiv|+^{N_f} \rangle$. We will keep these choices in mind from now on. Notice that we still have $S=0$ at this stage. This region is summarized as the first of the three loci in the top leaf of Fig.~\ref{fig:hasse-largeNf}.

 For the second step, consider $H=\eta \cdot I$ as in Eq.~\ref{eq:flow-eI}. 
\begin{itemize}
    \item  $\eta=|-^{N_f-2}+^2\rangle$:  $\dot S=0$, and $\dot M\sim \langle +^{N_f-2}-^2| \gamma_{(2)} | +^{N_f}\rangle\sim  \dd z^{N_f-1} \wedge \dd z^{N_f}$. We find $\tilde I \sim \tau |+^{2k} -^{N_f-2k-2}+^2 \rangle$. Now $t(I,\tilde I)=2k+2$, consistent with $S=0$.
    \item $\eta=|-^{N_f}\rangle$: we find $\dot S\neq 0$, $\dot M \sim \langle +^{N_f}| \gamma_{(2)} | +^{N_f}\rangle\sim \sum_{i=1}^{N_f} \dd z^i \wedge \dd \bar z^{\bar i} $, leading to an $\tilde I $  of type $t=0$, consistent with $S\neq 0$. 
    \item $\eta=|+^2-^{N-2} \rangle$ gives $M = M_0 +\tau \dd z^1 \wedge \dd z^2$; with $\tau=-1$ we can lower the rank of $M$. $\tilde I=0$, so no $\tilde I$ is generated. Proceeding in this fashion we can reach the orbits $\{S=0, M^2=0, \tilde I=0\}$. Those with $\{S=0, M^2=0, I=0\}$ can be reached similarly.   
\end{itemize}
Overall, we see that by changing $\eta$ we can reach orbits of type $0$ (with $S\neq 0$) and of type $\geq 2(k+1)$ ($S=0$); we summarize these two as the two remaining blocks on the top leaf of Fig.~\ref{fig:hasse-largeNf}. These are the only possibilities that are consistent with the relations, as we saw in Sec.~\ref{sec:pure}. 

For what concerns the slices, the procedure of iterated freezing of two-dimensional subspaces can be repeated until we reach overlap $2k-2$. The penultimate slice corresponds necessarily, due to our selection rules, to a zero-dimensional overlap. The corresponding slice is of exceptional type and is determined by the residual directions in the annihilators; the moduli space of the residual disjoint annihilators corresponds to $e_{N_f-2k+3}$, as there are $N_f-2(k-1)$ residual directions in each annihilator.\\

\subsection{Mass of the instanton states}
We will now provide a physical consequence of the described stratification, involving the mass of instanton states.
The supersymmetric algebra with eight supercharges in five dimensions admits  a central charge:
\begin{equation}
    \{Q^i_{\alpha},Q^j_{\beta}\}=\epsilon^{ij}(\Gamma^{\mu}C)_{[\alpha\beta]}P_{\mu}+\ii\epsilon^{ij}C_{[\alpha\beta]}Z+(\Gamma^{\mu\nu}C)_{(\alpha\beta)}Z_{\mu\nu}^{(ij)
    }
\end{equation}
with $i,j=1,2$ fundamental $SU(2)_R$ indices, $\alpha,\beta=1,\dots 4$ indices of the symplectic Majorana spinors, $C_{\alpha\beta}$ the charge conjugation matrix. BPS particles have mass, or tension, set by the central charge:
\begin{equation}
    M=|Z|=\left|\vec{q} \cdot \vec{\phi}+I \dfrac{1}{g^2}+\sum_{i=1}^{N_f}n_i m_i \right|
    \label{eq:BPSmass}
\end{equation}
with $\vec{q}$ representing the charges of the BPS state under the gauge symmetry, $\{\phi_i\}_{i=1\dots k}$ moduli parameterising the Coulomb branch. Furthermore, $g$ and $m_i$ are parameters classifying families of Coulomb branch, namely scalars in background vector multiplets. Moreover, $\vec{n}$ are the weights of the representation of the state under flavour symmetries and $I$ is the topological charge. For quiver theories, there are in general several topological charges.\\
At infinite coupling, we can tune the parameters $m_i$ to zero, corresponding to the masses of the flavours in the braneweb. Furthermore, the contribution with $g^{-2}$ vanishes and we are left solely with $\phi$. Indeed, even though there is no notion of gauge symmetry at strong coupling, we can still have Coulomb and mixed branches.\\

 It can appear nonsensical to consider nonzero charges $\vec{q}$, multiplying the Coulomb branch parameters, since we assumed the instanton operator to be gauge invariant on the Higgs branch. However, from a braneweb construction \cite{aharony1997branes, aharony1998webs}, the instanton state corresponds to a D1 string stretched between NS5 branes, whose position does depend on the vev of the scalar in the vector multiplet. Hence the instanton state is a gauge dependent state on the Coulomb branch. \\

 At infinite coupling, the Coulomb branch is still $k$-dimensional and opens up new directions at given symplectic leaves. To understand exactly when this happens, let us recall that the instantons are massless on a generic point on the Higgs branch, i.e.~on the largest leaf. However, at the penultimate leaf and below, only mesons of various ranks parameterise the Higgs branch, signaling a decoupling of the instantons from the chiral ring. This decoupling may be due to the instantons acquiring a mass, as we assumed in \cite{hanany2025chiral}. Hence, at the penultimate leaf, one of the $\phi$ components in Eq.~\ref{eq:BPSmass} is nonzero, opening up a direction of the Coulomb branch. In Eq.~\ref{eq:BPSmass}, no more than $k$ independent components of $\phi$ can be switched on, corresponding to the $k$ leaves below the top leaf. Each leaf corresponds to a change in mass of the instantons or, from the argument above, a change in the overlap of the annihilators. At a given leaf $m$, counting from the origin starting from zero, the number of non-vanishing components of $\phi$ is $k-m$.\\

 Hence, while the physical consequence of the stratification at finite coupling is the subsequent Higgsing of the gauge group, at infinite coupling the different symplectic leaves correspond to varying mass of the instanton states. We name this phenomenon \textit{instanton magnetization}, as on higher leaves the instantons have disjoint annihilator and are massless, while on lower leaves they tend to align and acquire a mass.

\section{Conclusions}
We set out this paper by asking what the meaning is behind the chiral ring relations in Tab.~\ref{tab:infinitecouplingconstraintsGENERAL} and whether we can find the Higgs branch stratification, for the theories at hand, independently of quiver subtraction.
We have found that not all chiral ring relations are independent: it is indeed sufficient to assume purity of $I$ and $\tilde{I}$, together with Eq.~\ref{eq:II} for $q=0,2$, and all other relations are merely a consequence. Furthermore, unexpected relations are also inferred at higher R-charge. Hence, the only independent degrees of freedom at strong coupling are the instanton and the anti-instanton; the mesons and gaugino bilinear at weak coupling are simply composites thereof.\\

We have bootstrapped the Poisson brackets of the chiral ring generators and we have explained how, from the Liouville-Arnold theorem, we can describe the Higgs branch at hand as an integrable system. The special role taken by the instantons descends from their being the action variables on the base of the Lagrangian fibration. From an algebraic point of view, the HWG of the holomorphic functions on the Higgs branch can indeed be expanded into instanton sectors, with a universal dressing. One question we can ask is, when the HWG can be expanded in this way, whether the dressed instantons parameterize a Lagrangian submanifold for the other gauge groups as well. \\

Considering the orbits of the bispinor defined by the product of the two pure spinors, we showed how to recover the stratification of the Higgs branch, which corresponds to the degeneration of the fibration upon the would-be union of two cones. We interpret the stratification in terms of the mass of the instanton themselves. While on the top leaf they are massless, lower leaves correspond to openings of mixed Coulomb-Higgs branches, where the mass of the instanton varies. Since they acquire a mass, they decouple from the chiral ring. Since the stratification is a consequence of the alignment of the spinors in flavour space and their becoming massive, we have called this phenomenon instanton magnetization.\\

It should be noted that the analysis carried out in this paper is valid for the combination of colors and flavors without symmetry enhancement, namely $N_f \leq 2k+3$. For the remaining two cases $N_f\in \{2k+4,2k+5\}$, while the chiral ring are known, there is no constraint requiring purity of the spinors. The cases with symmetry enhancement are problematic for other gauge groups as well, where it is not always guaranteed that the HWGs can be expanded in dressed instanton sectors. Another point concerns the last slice of Fig.~\ref{fig:hasse-Nfsmall}. We were not able to recover the order of the Klein singularity; indeed, in this diagram, the last slice is the only place where $k$ appears. This remains to be explained.\\

On a perhaps related note, we observe that the number of leaves of the Hasse diagram is always at most $k+1$, which corresponds to the R-charge of the instantons. It is not known why the R-charge should bound the number of phases; while the quantum numbers of the instantons are well defined and constant along every leaf, they are allowed to vary along the slices. It is then possible that a bound on the number of phases is tantamount to some condition on the operators populating the slices. This hypothesis will need to be explored in future work and extended to other gauge groups, as well as products of gauge groups.

\subsection*{Acknowledgments}
We would like to thank R.~Argurio, S.~Bennett, A.~Collinucci, M.~de Marco and C.~Hull for interesting questions and comments. The work of A.H.~and E.VdD.~is partially supported by STFC Consolidated Grant ST/X000575/1. A.T.~is supported in part by the INFN, and by the MUR-PRIN contract 2022YZ5BA2.

\bibliographystyle{JHEP}
\bibliography{bibli}

@article{tachikawa2015instanton,
  title={Instanton operators and symmetry enhancement in 5d supersymmetric gauge theories},
  author={Tachikawa, Yuji},
  journal={Progress of Theoretical and Experimental Physics},
  volume={2015},
  number={4},
  pages={043B06},
  year={2015},
  publisher={Oxford University Press},
  eprint={1501.01031},
  archivePrefix={arXiv},
  primaryClass={hep-th}
}

@article{hanany2025higgs,
  title={Higgs branch of 5d $\mathcal{N}=1$ symplectic gauge theories and dressed instanton operators},
  author={Hanany, Amihay and Driessche, Elias Van den},
  year={2025},
  eprint={2507.08669},
  archivePrefix={arXiv},
  primaryClass={hep-th}
}

@article{hanany2025chiral,
  title={Chiral ring along the {RG} flow in 5d $\mathcal{N}=1$},
  author={Hanany, Amihay and Driessche, Elias Van den},
  year={2025},
  eprint={2510.15635},
  archivePrefix={arXiv},
  primaryClass={hep-th}
}

@article{nekrasov2005lectures,
  title={Lectures on curved beta-gamma systems, pure spinors, and anomalies},
  author={Nekrasov, Nikita},
  year={2005},
  eprint={hep-th/0511008},
  archivePrefix={arXiv},
  primaryClass={hep-th}
}

@article{cremonesi2014coulomb,
  title={Coulomb branch and the moduli space of instantons},
  author={Cremonesi, Stefano and Ferlito, Giulia and Hanany, Amihay and Mekareeya, Noppadol},
  journal={Journal of High Energy Physics},
  volume={2014},
  number={12},
  pages={1--36},
  year={2014},
  publisher={Springer},
  eprint={1408.6835},
  archivePrefix={arXiv},
  primaryClass={hep-th}
}

@book{arnol2013mathematical,
  title={Mathematical methods of classical mechanics},
  author={Arnol'd, Vladimir Igorevich},
  volume={60},
  year={2013},
  publisher={Springer Science \& Business Media}
}

@inproceedings{nekrasov2009quantization,
  title={Quantization of integrable systems and four dimensional gauge theories},
  author={Nekrasov, Nikita A and Shatashvili, Samson L},
  booktitle={Proceedings, 16th international congress on mathematical physics (ICMP09): Prague, Czech Republic},
  pages={265--289},
  year={2009},
  organization={World Scientific},
  eprint={0908.4052},
  archivePrefix={arXiv},
  primaryClass={hep-th}
}

@article{donagi1996supersymmetric,
  title={Supersymmetric {Yang--Mills} theory and integrable systems},
  author={Donagi, Ron and Witten, Edward},
  journal={Nuclear Physics B},
  volume={460},
  number={2},
  pages={299--334},
  year={1996},
  publisher={Elsevier},
  eprint={hep-th/9510101},
  archivePrefix={arXiv},
  primaryClass={hep-th}
}

@article{cederwall2012geometry,
  title={The geometry of pure spinor space},
  author={Cederwall, Martin},
  journal={Journal of High Energy Physics},
  volume={2012},
  number={1},
  pages={1--11},
  year={2012},
  publisher={Springer},
  eprint={1111.1932},
  archivePrefix={arXiv},
  primaryClass={hep-th}
}

@article{gledhill2023poisson,
  title={Poisson brackets for some {Coulomb} branches},
  author={Gledhill, Kirsty and Hanany, Amihay},
  journal={Journal of High Energy Physics},
  volume={2023},
  number={3},
  pages={1--56},
  year={2023},
  publisher={Springer},
  eprint={2210.02966},
  archivePrefix={arXiv},
  primaryClass={hep-th}
}

@article{bourget2020higgs,
  title={The {Higgs} mechanism --- {Hasse} diagrams for symplectic singularities},
  author={Bourget, Antoine and Cabrera, Santiago and Grimminger, Julius F and Hanany, Amihay and Sperling, Marcus and Zajac, Anton and Zhong, Zhenghao},
  journal={Journal of High Energy Physics},
  volume={2020},
  number={1},
  pages={1--67},
  year={2020},
  publisher={Springer},
  eprint={1908.04245},
  archivePrefix={arXiv},
  primaryClass={hep-th}
}

@incollection{donagi1995spectral,
  title={Spectral covers, algebraically completely integrable, Hamiltonian systems, and moduli of bundles},
  author={Donagi, Ron and Markman, Eyal},
  booktitle={Integrable Systems and Quantum Groups: Lectures given at the 1st Session of the Centro Internazionale Matematico Estivo (CIME) held in Montecatini Terme, Italy, June 14--22, 1993},
  pages={1--119},
  year={2006},
  publisher={Springer},
   eprint={alg-geom/9507017} 
}

@article{bourget2020magnetic,
  title={Magnetic quivers from brane webs with {O5} planes},
  author={Bourget, Antoine and Grimminger, Julius F and Hanany, Amihay and Sperling, Marcus and Zhong, Zhenghao},
  journal={Journal of High Energy Physics},
  volume={2020},
  number={7},
  pages={1--82},
  year={2020},
  publisher={Springer},
  eprint={2004.04082},
  archivePrefix={arXiv},
  primaryClass={hep-th}
}

@article{cabrera2018quiver,
  title={Quiver subtractions},
  author={Cabrera, Santiago and Hanany, Amihay},
  journal={Journal of High Energy Physics},
  volume={2018},
  number={9},
  pages={1--21},
  year={2018},
  publisher={Springer},
  eprint={1803.11205},
  archivePrefix={arXiv},
  primaryClass={hep-th}
}

@inproceedings{beem2020secondary,
  title={Secondary products in supersymmetric field theory},
  author={Beem, Christopher and Ben-Zvi, David and Bullimore, Mathew and Dimofte, Tudor and Neitzke, Andrew},
  booktitle={Annales Henri Poincare},
  volume={21},
  pages={1235--1310},
  year={2020},
  organization={Springer},
  eprint={1809.00009},
  archivePrefix={arXiv},
  primaryClass={hep-th}
}

@article{hanany2014highest,
  title={Highest weight generating functions for {Hilbert} series},
  author={Hanany, Amihay and Kalveks, Rudolph},
  journal={Journal of High Energy Physics},
  volume={2014},
  number={10},
  pages={1--68},
  year={2014},
  publisher={Springer},
  eprint={1408.4690},
  archivePrefix={arXiv},
  primaryClass={hep-th}
}

@article{beem2017deformation,
  title={Deformation quantization and superconformal symmetry in three dimensions},
  author={Beem, Christopher and Peelaers, Wolfger and Rastelli, Leonardo},
  journal={Communications in Mathematical Physics},
  volume={354},
  number={1},
  pages={345--392},
  year={2017},
  publisher={Springer},
  eprint={1601.05378},
  archivePrefix={arXiv},
  primaryClass={hep-th}
}

@book{tomasiello2022geometry,
  title={Geometry of string theory compactifications},
  author={Tomasiello, Alessandro},
  year={2022},
  publisher={Cambridge University Press}
}

@article{Bennett:2024loi,
    author = "Bennett, Sam and Hanany, Amihay and Kumaran, Guhesh and Li, Chunhao and Liu, Deshuo and Sperling, Marcus",
    title = "{Quiver subtraction on the {Higgs} branch}",
    eprint = "2409.16356",
    archivePrefix = "arXiv",
    primaryClass = "hep-th",
    reportNumber = "Imperial/TP/24/AH/03, Imperial/TP/24/AH/03",
    doi = "10.1016/j.nuclphysb.2025.116917",
    journal = "Nucl. Phys. B",
    volume = "1016",
    pages = "116917",
    year = "2025"
}

@article{Kim:2012gu,
    author = "Kim, Hee-Cheol and Kim, Sung-Soo and Lee, Kimyeong",
    title = "{5-dim Superconformal Index with Enhanced {$E_n$} Global Symmetry}",
    eprint = "1206.6781",
    archivePrefix = "arXiv",
    primaryClass = "hep-th",
    reportNumber = "KIAS-P12033",
    doi = "10.1007/JHEP10(2012)142",
    journal = "JHEP",
    volume = "10",
    pages = "142",
    year = "2012"
}

@article{bennett2026quotient,
  title={Quotient Quiver Subtraction ---Classical Groups},
  author={Bennett, Sam and Hanany, Amihay and Kumaran, Guhesh},
  year={2026},
  eprint={2603.08774},
  archivePrefix={arXiv},
  primaryClass={hep-th}
}

@article{aharony1998webs,
  title={Webs of $(p,q)$ 5-branes, five dimensional field theories and grid diagrams},
  author={Aharony, Ofer and Hanany, Amihay and Kol, Barak},
  journal={Journal of High Energy Physics},
  volume={1998},
  number={01},
  pages={002},
  year={1998},
  publisher={IOP Publishing},
  eprint={hep-th/9710116},
  archivePrefix={arXiv},
  primaryClass={hep-th}
}

@article{cabrera2019tropical,
  title={Tropical geometry and five dimensional {Higgs} branches at infinite coupling},
  author={Cabrera, Santiago and Hanany, Amihay and Yagi, Futoshi},
  journal={Journal of High Energy Physics},
  volume={2019},
  number={1},
  pages={1--50},
  year={2019},
  publisher={Springer},
  eprint={1810.01379},
  archivePrefix={arXiv},
  primaryClass={hep-th}
}

@article{aharony1997branes,
  title={Branes, superpotentials and superconformal fixed points},
  author={Aharony, Ofer and Hanany, Amihay},
  journal={Nuclear Physics B},
  volume={504},
  number={1-2},
  pages={239--271},
  year={1997},
  publisher={Elsevier},
  eprint={hep-th/9704170},
  archivePrefix={arXiv},
  primaryClass={hep-th}
}

@article{seiberg1996five,
  title={Five dimensional {SUSY} field theories, non-trivial fixed points and string dynamics},
  author={Seiberg, Nathan},
  journal={Physics Letters B},
  volume={388},
  number={4},
  pages={753--760},
  year={1996},
  publisher={Elsevier},
  eprint={hep-th/9608111},
  archivePrefix={arXiv},
  primaryClass={hep-th}
}

@article{zafrir2015instanton,
  title={Instanton operators and symmetry enhancement in 5d supersymmetric {USp}, {SO} and exceptional gauge theories},
  author={Zafrir, Gabi},
  journal={Journal of High Energy Physics},
  volume={2015},
  number={7},
  pages={1--34},
  year={2015},
  publisher={Springer},
  eprint={1503.08136},
  archivePrefix={arXiv},
  primaryClass={hep-th}
}

@book{collingwood1993nilpotent,
  title={Nilpotent orbits in semisimple {Lie} algebra: an introduction},
  author={Collingwood, David H and McGovern, William M},
  year={1993},
  publisher={CRC Press}
}

@article{cremonesi2017instanton,
  title={Instanton operators and the {Higgs} branch at infinite coupling},
  author={Cremonesi, Stefano and Ferlito, Giulia and Hanany, Amihay and Mekareeya, Noppadol},
  journal={Journal of High Energy Physics},
  volume={2017},
  number={4},
  pages={1--43},
  year={2017},
  publisher={Springer},
  eprint={1505.06302},
  archivePrefix={arXiv},
  primaryClass={hep-th}
}

@article{Beauville:1999jhe,
    author = "Beauville, Arnaud",
    title = "{Symplectic singularities}",
    eprint = "math/9903070",
    archivePrefix = "arXiv",
    doi = "10.1007/s002229900043",
    journal = "Invent. Math.",
    volume = "139",
    number = "3",
    pages = "541--549",
    year = "2000"
}

@article{Kaledin2006,
  author  = {Kaledin, Dmitry},
  title   = {Symplectic singularities from the {Poisson} point of view},
 eprint = "math/0310186", 
  journal = {Journal f{\"u}r die reine und angewandte Mathematik (Crelles Journal)},
  year    = {2006},
  volume  = {2006},
  number  = {600},
  pages   = {135--156},
  doi     = {10.1515/CRELLE.2006.089}
}

@article{Namikawa2001,
  author  = {Namikawa, Yoshinori},
  title   = {Extension of 2-forms and symplectic varieties},
   eprint = "math/0010114",
  journal = {Journal f{\"u}r die reine und angewandte Mathematik (Crelles Journal)},
  year    = {2001},
  volume  = {2001},
  number  = {539},
  pages   = {123--147},
  doi     = {https://doi.org/10.1515/crll.2001.070}
}
\end{document}